\documentclass[aps,prc,nofootinbib,superscriptaddress,twocolumn,10pt,floatfix]{revtex4-2}

\usepackage{emnorm2}
\usepackage{writelabel}

\ifproofpre{}{\ifeof18
    \typeout{Version control information not updated!}
\else
    \immediate\write18{vc/vc}
\fi

\RequirePackage{xcolor}

\IfFileExists{vc/vc-info.tex}{\gdef\GITAbrHash{f751bf5}\gdef\VCRevision{\GITAbrHash}\gdef\VCDateText{July 22, 2025}\gdef\VCRevisionMod{\VCRevision}\gdef\VCRevisionMod{\VCRevision} }{
\gdef\GITAbrHash{unknown}\gdef\VCRevision{\GITAbrHash}\gdef\VCDateText{\today}\gdef\VCRevisionMod{\VCRevision}}
 }

\begin{document}
\ifproofpre{}{\count\footins = 1000}

\title{Robust \textit{ab initio} predictions for dimensionless ratios of $E2$ and radius observables. I.~Electric quadrupole moments and deformation}

\author{Mark A.~Caprio\,\orcidlink{0000-0001-5138-3740}}
\affiliation{Department of Physics and Astronomy, University of Notre Dame, Notre Dame, Indiana 46556-5670, USA}

\author{Pieter Maris\,\orcidlink{0000-0002-1351-7098}}
\affiliation{Department of Physics and Astronomy, Iowa State University, Ames, Iowa 50011-3160, USA}

\author{Patrick J.~Fasano\,\orcidlink{0000-0003-2457-4976}}
\altaffiliation[Present address: ]{NextSilicon Inc., Minneapolis, Minnesota 55402-1572, USA}
\affiliation{Department of Physics and Astronomy, University of Notre Dame, Notre Dame, Indiana 46556-5670, USA}
\affiliation{Physics Division, Argonne National Laboratory, Argonne, Illinois 60439-4801, USA}

\date{\ifproofpre{\today}{\VCDateText}}

\begin{abstract}
  Converged results for $E2$ observables are notoriously challenging to obtain in
\textit{ab initio} no-core configuration interaction (NCCI) approaches.  Matrix
elements of the $E2$ operator are sensitive to the large-distance tails of the
nuclear wave function, which converge slowly in an oscillator basis expansion.
Similar convergence challenges beset \textit{ab initio} prediction of the
nuclear charge radius.  However, we exploit systematic correlations between the
calculated $E2$ and radius observables to yield meaningful predictions for
relations among these observables.  In particular, we examine \textit{ab initio}
predictions for dimensionless ratios of the form $Q/r^2$, for nuclei throughout
the $p$ shell.  Meaningful predictions for electric quadrupole moments may then be made by
calibrating to the ground-state charge radius, if experimentally known,
or \textit{vice versa}.  Moreover, these dimensionless ratios provide \textit{ab
initio} insight into the nuclear quadrupole deformation.
 \end{abstract}

\ifproofpre{}{\preprint{Git hash: \VCRevisionMod}}

\maketitle

\writelabel{part2:fig:be2-norm-rp-scan-12c}{7}

\section{Introduction}
\label{sec:intro}

Converged \textit{ab initio} predictions for nuclear electric quadrupole ($E2$)
observables, including transition strengths and moments, are challenging to
obtain~\cite{pervin2007:qmc-matrix-elements-a6-7,bogner2008:ncsm-converg-2N,maris2013:ncsm-pshell,carlson2015:qmc-nuclear,odell2016:ir-extrap-quadrupole,roth2023:em-properties-nuclei}.
\textit{Ab initio} no-core configuration interaction (NCCI), or no-core
shell-model (NCSM), calculations~\cite{barrett2013:ncsm} rely upon a Slater
determinant expansion of the wave function (conventionally in an oscillator
basis) and must, in practice, be carried out in a finite, truncated basis.  The
results are only approximations to the true values which would be obtained
by solving the full, untruncated many-body problem.  Long-range observables,
\textit{i.e.}, those which are sensitive to the large-distance tails of the
nuclear wave function, such as $E2$ matrix elements, are slowly convergent in
NCCI calculations, as these tails are described only with difficulty in an
oscillator-basis expansion.  Including a sufficiently large basis to obtain
meaningful predictions often becomes computationally prohibitive.

Nonetheless, one may exploit systematic correlations among calculated
observables to yield meaningful predictions for relations among these
observables, even where the observables individually are not adequately
converged.  The convergence patterns of calculated $E2$ matrix elements may be
strongly correlated~\cite{calci2016:observable-correlations-chiral}, especially
for $E2$ matrix elements among states sharing similar structure, \textit{e.g.},
members of low-lying rotational
bands~\cite{caprio2013:berotor,maris2015:berotor2,*maris2019:berotor2-ERRATUM,calci2016:observable-correlations-chiral,caprio2019:bebands-sdanca19,caprio2020:bebands}
or mirror states~\cite{henderson2019:7be-coulex,caprio2021:emratio}.  Thus, most
naturally, if a single $E2$ observable is well-known from experiment, a
meaningful prediction may then be made for the other, correlated matrix
elements~\cite{calci2016:observable-correlations-chiral}.

In particular, the ground-state quadrupole moment is precisely measured in many
nuclei~\cite{stone2016:e2-moments}, as summarized for $p$-shell nuclei in
Fig.~\ref{fig:nuclear-chart}.  In these cases, the known ground-state quadrupole
moment provides a calibration reference, permitting $E2$ transition strengths to
be estimated based on robust \textit{ab initio} NCCI predictions for
$B(E2)/(eQ)^2$.  This approach is illustrated for various light $p$-shell nuclei
($A\leq9$) in Ref.~\cite{caprio2022:8li-trans}.  Predictions for $B(E2)/(eQ)^2$
are also considered for several $\isotope{C}$ isotopes (and $\isotope[10]{Be}$)
in Ref.~\cite{li2024:quadrupole-10be-c}.

\begin{figure}[b]
\begin{center}
\includegraphics[width=\ifproofpre{1}{1}\hsize]{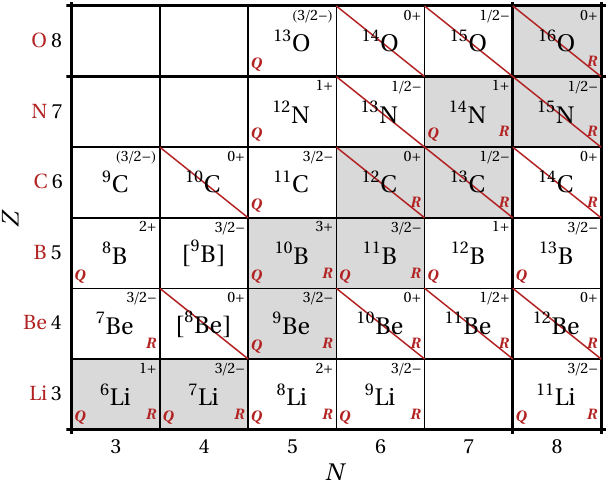}
\end{center}
\caption{Overview of particle-bound nuclides in the $p$ shell, where those with
  measured ground-state quadrupole moments~\cite{stone2016:e2-moments} and
  charge radii~\cite{angeli2013:charge-radii,npa2017:012} are indicated by the
  letter ``$Q$'' or ``$R$'', respectively.  Brackets indicate a particle-unbound
  but narrow ($\lesssim 1\,\keV$) ground-state resonance, shading indicates
  beta-stable nuclides, and the experimental ground-state angular momentum and
  parity are
  given~\cite{npa2002:005-007,npa2004:008-010,npa2012:011,npa2017:012,npa1991:013-015}.
  Nuclei for which the ground-state angular momentum does not support a
  quadrupole moment ($J\leq1/2$) are crossed out with a diagonal line.  }
\label{fig:nuclear-chart}
\end{figure}

However, calibration to the ground-state quadrupole moment is subject to the
limitation that only states with angular momentum $J\geq1$ admit a nonvanishing
quadrupole moment.  Thus, notably, calibration to the ground-state
quadrupole moment is not possible for even-even nuclei, nor for odd-mass nuclei
with $J=1/2$ ground states.  (Nuclei not supporting a ground-state quadrupole
moment are crossed out with diagonal line in Fig.~\ref{fig:nuclear-chart}.)

We observe that the convergence of calculated $E2$ matrix elements may also be
correlated with the convergence of the calculated electric monopole ($E0$)
moment or, equivalently, root mean square (r.m.s.) radius.  The known
ground-state charge radius is, like the quadrupole moment, precisely measured in
many nuclei~\cite{angeli2013:charge-radii} (again, as summarized for $p$-shell
nuclei in Fig.~\ref{fig:nuclear-chart}).  Thus, the measured ground-state charge
radius, like the quadrupole moment, may be used as a calibration reference,
permitting $E2$ transition strengths to be estimated based on robust \textit{ab
  initio} NCCI predictions for ratios of the form $B(E2)/(e^2r^4)$, where $r$ is
the r.m.s.\ radius, without being subject to any constraint on the ground-state
angular momentum.  However, of course, the determination of either the
ground-state quadrupole moment or radius is subject to practical experimental
considerations~\cite{neugart2006:nuclear-moments,yang2023:laser-spectroscopy-nuclei},
including typically that the nucleus be particle bound.

Similarly, predictions for ratios of the form $Q/r^2$ permit estimation of the
ground-state (or an excited-state) quadrupole moment from a measured radius, or
\textit{vice versa}.  Only for $\isotope[7]{Be}$, among $p$-shell nuclei, is the
radius known but quadrupole moment unknown.  However, for the proton-rich
nuclides $\isotope[8]{B}$, $\isotope[11]{C}$, and $\isotope[12]{N}$, as well as
neutron-rich $\isotope[12]{B}$, the quadrupole moment is known while the radius
is unknown, as seen from Fig.~\ref{fig:nuclear-chart}.

Such ratios $Q/r^2$ also provide \textit{ab initio} insight into the quadrupole
deformation.  Namely, in an axially symmetric rotational picture, they provide a
measure of the Bohr deformation variable $\beta$~\cite{bohr1952:vibcoupling} or,
more precisely, its microscopically defined counterpart obtained through the
quadrupole tensor~\cite{rowe1985:micro-collective-sp6r}.  We may consider such
ratios not only for proton observables $Q_p$ and $r_p$, providing a measure of
the deformation $\beta_p$ for the proton distribution, but also for neutron
observables $Q_n$ and $r_n$, providing a measure of the deformation $\beta_n$ of
the neutron distribution, which may in general be distinct.  Although the proton
observables are more readily accessible to experiment, through electromagnetic
probes, reaction observables are sensitive to the neutron observables
(\textit{e.g.},
Refs.~\cite{bernstein1983:pn-me-hadron-scatt,jonson2004:light-dripline,tanihata2013:halo-expt,furuno2019:10c-alphaalphaprime-neutron-quadrupole}),
at least in a model-dependent fashion.  Moreover, parity nonconservation effects
in atomic or molecular systems are sensitive to the nuclear weak charge
distribution, and thus can provide an alternative means of probing these neutron
observables~\cite{flambaum2017:quadrupole-moment-parity-nonconservation,arrowsmithkron2024:radioactive-molecules}.

In the present article (Part~I), we focus on the \textit{ab initio} prediction
of $Q/r^2$, while calibration of \textit{ab initio} predictions for $E2$
strengths to the charge radius is the subject of a subsequent article
(Part~II)~\cite{emnorm2-part2}.  We first lay out the expected relations between
$E2$ and radius observables, in terms of dimensionless ratios
(Sec.~\ref{sec:background:ratios}).  We also briefly review the conversion
between the experimentally accessible charge radius $r_c$ and the point-proton
radius $r_p$ that more naturally arises in nuclear structure calculations
(Sec.~\ref{sec:background:rc}) and comment on the physical interpretation of the
dimensionless ratio $Q/r^2$ in terms of the nuclear quadrupole deformation
(Sec.~\ref{sec:background:deformation}).  We then explore the convergence
obtained for the ratio, and compare the resulting predictions against
experiment.  Here we begin with $\isotope[9]{Be}$ as an illustrative case for
detailed exploration (Sec.~\ref{sec:results-moments:9be}), including an
examination of convergence diagnostics and a straightforward exponential basis
extrapolation (for which relations are provided in
Appendix~\ref{sec:app-geometric}).  We then consider the dimensionless ratio
$Q/r_p^2$ for the ground states of particle-bound nuclides across the $p$ shell
(Sec.~\ref{sec:results-moments:survey}).  We restrict attention to doubly
open-shell nuclei (\textit{i.e.}, with $3\leq Z,N \leq7$), since, in the
semimagic nuclei (at the $N,Z=8$ shell closure), mixing with intruder
configurations~\cite{jonson2004:light-dripline,heyde2011:shape-coexistence}
complicates the identification between calculated and physical ground states and
disrupts the ordinary convergence
patterns~\cite{caprio2021:emratio,mccoy2024:12be-shape}.  Finally, we translate
these results into an exploration of the evolution of the deformations of the
proton and neutron distributions within the nuclear ground state, under the
assumption of axial symmetry (Sec.~\ref{sec:results-moments:deformation}).
Preliminary results were reported in Ref.~\cite{caprio2022:emnorm}.

\section{Background: Dimensionless ratios of observables and their interpretation}
\label{sec:background}

\subsection{Dimensionless ratios}
\label{sec:background:ratios}

Electric quadrupole moments and $E2$ strengths both may be expressed\footnote{\label{fn:q-be2-rme}The quadrupole
  moment for a state of angular momentum $J$ is defined~\cite{suhonen2007:nucleons-nucleus} in terms of a stretched
  $E2$ matrix element, as
\begin{math}
  eQ(J)\equiv(16\pi/5)^{1/2}\tme{JJ}{Q_{20}}{JJ},
\end{math}
and thus may be written, by the Wigner-Eckart theorem~\cite{edmonds1960:am,varshalovich1988:am}, in terms of a reduced matrix element as
\begin{math}
  eQ(J)=(16\pi/5)^{1/2}(2J+1)^{-1/2}\tcg{J}{J}{2}{0}{J}{J}\trme{J}{Q_2}{J}.
\end{math}
The reduced transition probability
between states of angular momentum $J_i$ and $J_f$ is likewise expressed in terms of a reduced matrix element as
\begin{math}
  B(E2;J_i\rightarrow J_f)=(2J_i+1)^{-1}\abs{\trme{J_f}{Q_2}{J_i}}^2.
\end{math}
}
in terms of
reduced matrix elements of the $E2$ operator,
\begin{math}
Q_{2\mu}=\sum_{i\in p}e r_{i}^2Y_{2\mu}(\uvec{r}_{i}),
\end{math}
where the summation runs over the (charged)
protons.\footnote{\label{fn:e2-intrinsic}The $E2$ operator of interest is, more
precisely, defined thusly in terms of the \textit{intrinsic} coordinates,
taken relative to the nuclear center of mass frame~\cite{caprio2020:intrinsic}.}
Ratios of the form
\begin{equation}
  \label{eqn:ratio-rtp-q}
       \frac{B(E2)}{(eQ)^2} \propto \Biggl\lvert \frac{
         \trme{\cdots}{\,\sum_{i\in p}r_i^2Y_{2}(\uvec{r}_i)\,}{\cdots} }{
         \trme{\cdots}{\,\sum_{i\in p}r_i^2Y_{2}(\uvec{r}_i)\,}{\cdots} }
       \Biggr\rvert^2
\end{equation}
are dimensionless and involve like powers of $E2$ reduced matrix elements
(albeit involving different states) in the numerator and the denominator, much as in
ratios of quadrupole moments or in ratios of $E2$ strengths.  Truncation errors
may cancel in such ratios, as explored and exploited in
Refs.~\cite{calci2016:observable-correlations-chiral,caprio2022:8li-trans}.  In
this case, a known ground-state quadrupole moment provides a calibration
reference for the $E2$ strength.

The r.m.s.\ point-proton radius $r_p$ is evaluated in terms of the monopole
moment
\begin{math}
  M_0=\tme{J}{\sum_{i\in p} r_{i}^2}{J}
\end{math}
as
\begin{math}
  r_p=(M_0/Z)^{1/2}.
\end{math}  Thus,
$r_p^2$ is proportional to a matrix element of a one-body operator, the $E0$
operator, which once again involves an $r^2$ radial dependence.  Ratios
$Q/r_p^2$ or $B(E2)/(e^2r_p^4)$ involving an $E2$ observable and the appropriate
power of $r_p$ are again dimensionless, and involve ratios of similar powers of
matrix elements of operators sharing the same $r^2$ dependence.  Namely, these
ratios are of the form
\begin{equation}
  \label{eqn:ratio-q-r}
        \frac{Q}{r_p^2} \propto  \frac{ \trme{\cdots}{\,\sum_{i\in
        p}r_i^2Y_{2}(\uvec{r}_i)\,}{\cdots} }{ \trme{\cdots}{\,\sum_{i\in
                    p}r_i^2\,}{\cdots} }
\end{equation}
and
\begin{equation}
  \label{eqn:ratio-rtp-r}
      \frac{B(E2)}{(e^2r_p^4)} \propto \Biggl\lvert \frac{
        \trme{\cdots}{\,\sum_{i\in p}r_i^2Y_{2}(\uvec{r}_i)\,}{\cdots} }{
        \trme{\cdots}{\,\sum_{i\in p}r_i^2\,}{\cdots} } \Biggr\rvert^2,
\end{equation}
respectively.  It is not unreasonable to anticipate that errors might
again cancel in the ratio.  In this case, a known point-proton radius provides a
calibration reference for the $E2$ moment or strength.

\subsection{Relation of point-proton and charge radii}
\label{sec:background:rc}

To connect with experiment, we must relate the point-proton radius $r_p$
appearing in nuclear structure calculations to the charge radius $r_c$ probed in
experiment.  At leading order in the inverse nucleon mass, the relationship is
expressed as a set of additive corrections to
$r_p^2$~\cite{friar1975:nuclear-charge-distributions,friar1977:pion-exchange,friar1980:deuteron-charge,friar1997:charge-radius-correction,lu2013:laser-neutron-rich}:
\begin{equation}
  \label{eqn:rc}
  r_c^2=r_p^2+\Bigl(\frac{3}{4M_p^2}+R_p^2\Bigr)+\frac{N}{Z}R_n^2 + r^2_{\text{s.o.}}
  + r^2_{\text{m.e.c.}},
\end{equation}
where the meanings of the terms on the right hand side are discussed below.

The first two correction terms relative to $r_p^2$, in~(\ref{eqn:rc}), reflect
the charge distributions within the nucleons themselves, and are independent of
nuclear structure.  The term in parentheses accounts for the mean-square charge
radius of the proton, which is customarily separated into a Darwin-Foldy
contribution [$3/(4M_p^2)=0.033\,\fm^2$, where $M_p$ is the proton mass],
representing the finite charge radius which would be generated due to
relativistic effects even for a point proton, and a remaining contribution,
taken to represent the effect of the non-pointlike hadronic structure of the
proton.  It is this latter contribution which is conventionally tabulated as the
mean-square charge radius of the proton (denoted in the hadronic physics
literature again by $r_p^2$~\cite{pdg2022}, but renamed in the nuclear physics
literature, \textit{e.g.}, to $R_p^2$~\cite{lu2013:laser-neutron-rich}, to avoid
overloading the use of $r_p$ in the nuclear sense as defined above in
Sec.~\ref{sec:background:ratios}).  The $R_n^2$ term similarly accounts for the
mean-square charge radius of the neutron, which reflects the inhomogenous charge
distribution arising within the (overall neutral) neutron.  In the present work,
we use $R_p=0.8409(4)\,\fm$ [giving $R_p^2+3/(4M_p^2)=0.7401(7)\,\fm^2$] and
$R_n^2=-0.1155(17)\,\fm^2$~\cite{pdg2022}.

The remaining two terms represent the relativistic spin-orbit effect (which may
be interpreted as a contribution to the charge density arising from Lorentz
transformation of the nucleon anomalous magnetic moments) and meson exchange
contributions.  Their contributions depend in detail upon the nuclear structure
and, in the latter case, upon the assumed form of the meson exchange
contributions to the charge
operator~\cite{pastore2011:em-charge-n4lo,kolling2011:em2b,krebs2020:nuclear-currents}.
(See Ref.~\cite{ong2010:charge-radius-correction-halo} for estimates of the
spin-orbit correction.)  The corrections in their entirety may in principle be
obtained from a chiral effective field theory ($\chi$EFT) expansion, yielding
two-body and higher contributions.  When extracting values for $r_p$ from the
measured $r_c$~\cite{angeli2013:charge-radii}, in the present work, we only
account for the first two correction terms in~(\ref{eqn:rc}), while neglecting
the remaining (structure-dependent and less well-constrained) terms.
Nonetheless, it is important to keep in mind their possible relevance.

\subsection{Relation to deformation}
\label{sec:background:deformation}

The dimensionless ratios considered above, in Sec.~\ref{sec:background:ratios},
are of interest not only for their prospective convergence properties, but also
for their physical significance, in the context of rotation and deformation.
Ratios of $E2$ matrix elements are a direct prediction of rotational models and
depend only upon the nature of the rotation (\textit{e.g.}, axial or triaxial),
not directly on the overall magnitude of the deformation.
For an
axially symmetric rotor~\cite{rowe2010:collective-motion}, all $E2$ matrix elements within a rotational band
are proportional to the same intrinsic quadrupole moment
\begin{equation}
  \label{eqn:Q0-as-me}
  eQ_0\equiv\Bigl(\frac{16\pi}{5}\Bigr)^{1/2}\tme{\phi_K}{Q_{2,0}}{\phi_K},
\end{equation}
and the ratio $B(E2)/(eQ)^2$, of~(\ref{eqn:ratio-rtp-q}), for a transition
strength and quadrupole moment within the same rotational band, with rotational intrinsic state $\tket{\phi_K}$, is given simply in terms of Clebsch-Gordan
coefficients.\footnote{\label{fn:rot-reln}For an axially symmetric rotor, within
a rotational band with angular momentum projection $K$ along the intrinsic
symmetry axis, and rotational intrinsic state $\tket{\phi_K}$, $E2$ reduced
matrix elements are given in terms of the intrinsic quadrupole moment $Q_0$ by
\begin{math}
\label{eqn:rotational-e2-rme}
\trme{KJ_f}{Q_2}{KJ_i}
= \bigl[5/(16\pi)\bigr]^{1/2}\hat{J}_i \tcg{J_i}{K}{2}{0}{J_f}{K} e Q_0,
\end{math}
where $\hat{J}\equiv(2J+1)^{1/2}$, although it should be noted that an
additional cross term may contribute for $1/2\leq K \leq
1$~\cite{rowe2010:collective-motion} (see, \textit{e.g.}, Sec.~II\,C of
Ref.~\cite{maris2015:berotor2} for reference plots and a summary of the relevant
relations).
For the spectroscopic quadrupole moment
(footnote~\ref{fn:q-be2-rme}), the rotational relation becomes
\begin{math}
Q(J)=[3K^2-J(J+1)]Q_0/[(J+1)(2J+3)].
\end{math}
}  In contrast, the ratio of an $E2$ matrix element to the monopole
moment depends upon the structure of the rotational intrinsic state and provides
a measure of the overall magnitude of the deformation. In particular, the
axially symmetric rotational picture directly relates the ratios $Q/r_p^2$ and
$B(E2)/(e^2r_p^4)$, appearing in~(\ref{eqn:ratio-q-r})
and~(\ref{eqn:ratio-rtp-r}), to the $\beta$ deformation of the rotational
intrinsic state.

The Bohr deformation variable $\beta$~\cite{bohr1952:vibcoupling} is traditionally introduced in a purely
macroscopic context.  In terms of the quadrupole ($\lambda=2$) surface deformation
parameters $\alpha_{2\mu}$ appearing in the multipole expansion
\begin{math}
  R(\theta,\phi)=R_0\bigl[1+\sum_{\lambda\mu}\alpha_{\lambda\mu}^*Y_{\lambda\mu}(\theta,\phi)\bigr]
\end{math}
for the surface displacement of a deformed liquid drop, $\beta$ is defined as
the spherical tensor norm of $\alpha_{2}$,
via~\cite{bohr1952:vibcoupling,eisenberg1987:v1-SERIES,bohr1998:v2}
\begin{equation}
  \label{eqn:beta-as-norm}
  \beta^2\equiv\sum_\mu\abs{\alpha_{2\mu}}^2.
\end{equation}
This relation assumes a well-defined nuclear shape, but the deformation need not
be axially symmetric (\textit{i.e.}, it may be triaxial).

In microscopic nuclear theory, we work in terms of the quadrupole tensor, rather
than surface deformation parameters.  In making the connection to the liquid
drop picture, it is traditional to consider the \textit{matter} (or
point-nucleon) quadrupole tensor rather than the \textit{electromagnetic} (or
point-proton) quadrupole tensor of Sec.~\ref{sec:background:ratios}, and to work
in terms of the quadrupole moment operator $\Qmom_2$, normalized with a
conventional factor of
$(16\pi/5)^{1/2}$~\cite{loebner1970:intrinsic-quadrupole-moments-deformation,bohr1998:v1}.
Then, the integral expression for the quadrupole tensor in terms of the matter
density $\rho(\vec{r})$ is
\begin{math}
  \Qmom_{2\mu}=(16\pi/5)^{1/2}\int r^2 Y_{2\mu}(\uvec{r})\rho(\vec{r})\,d^3\vec{r},
\end{math}
and the microscopic representation (in terms of point nucleons) is
\begin{equation}
  \label{eqn:Qmom-defn}
\Qmom_{2\mu}=\Bigl(\frac{16\pi}{5}\Bigr)^{1/2}\sum_{i\in p,n} r_{i}^2Y_{2\mu}(\uvec{r}_{i}).
\end{equation}

In a liquid drop picture, it is straightforward to deduce a relation, valid at
leading order in the $\alpha_{2\mu}$~\cite{loebner1970:intrinsic-quadrupole-moments-deformation}, between the components $\Qmom_{2\mu}$ of
the quadrupole tensor and the deformation parameters $\alpha_{2\mu}$.  This
relation involves also the monopole moment (or radius).  Namely (see,
\textit{e.g.}, Sec.~3.1 of Ref.~\cite{rowe1985:micro-collective-sp6r}),
\begin{equation}
  \label{eqn:Q2mu-M0-beta-undifferentiated}
  \Qmom_{2\mu}=\Bigl(\frac{5}{\pi}\Bigr)^{1/2} M_0 \alpha_{2\mu}.
\end{equation}
Here, the \textit{matter} monopole moment
\begin{math}
  M_0=\int r^2 \rho(\vec{r})\,d^3\vec{r},
\end{math}
is representated in terms of point nucleons as
\begin{math}
  M_0=\sum_{i\in p,n} r_{i}^2.
\end{math}

The squared norm of the quadrupole tensor,
\begin{equation}
  \label{eqn:Q0-as-norm}
  \abs{Q_0}^2 \equiv \sum_\mu\abs{\Qmom_{2\mu}}^2,
\end{equation}
is rotationally invariant and thus, in a rotational picture, a property of the
rotational intrinsic state (and the same for all members of a rotational
band).  From the componentwise
relation~(\ref{eqn:Q2mu-M0-beta-undifferentiated}) between the tensors $\Qmom_2$
and $\alpha_2$, their norms $\abs{Q_0}$ and $\beta$ are similarly related as
\begin{equation}
  \label{eqn:Q0-M0-beta-undifferentiated}
  \abs{Q_0}=\Bigl(\frac{5}{\pi}\Bigr)^{1/2} M_0 \beta.
\end{equation}
We may now discard any reference to a classical liquid drop, and take the
relation~(\ref{eqn:Q0-M0-beta-undifferentiated}) to be a fully microscopic
defining expression for $\beta$.  Equivalently, in terms of the r.m.s.\ matter
radius $r$, defined by $M_0=Ar^2$,
\begin{equation}
  \label{eqn:Q0-r-beta-undifferentiated}
  \abs{Q_0}
  =\Bigl(\frac{5}{\pi}\Bigr)^{1/2} A r^2 \beta.
\end{equation}

Now let us return to electromagnetic (proton) observables, obtained with the
proton quadrupole operator of Sec.~\ref{sec:background:ratios}.  With
replacements $\abs{Q_0}\rightarrow \abs{Q_{0,p}}$, $r\rightarrow r_p$, $A\rightarrow Z$, and $\beta\rightarrow
\beta_p$ in~(\ref{eqn:Q0-r-beta-undifferentiated}), now restricting the
summation in~(\ref{eqn:Qmom-defn}) to protons, we obtain
\begin{equation}
  \label{eqn:Q0-r-beta-proton}
  \abs{Q_{0,p}}=\Bigl(\frac{5}{\pi}\Bigr)^{1/2} Z r_p^2 \beta_p.
\end{equation}
Similar relations may, of course, be written for neutron quadrupole observables,
with analogous replacements $\abs{Q_0}\rightarrow \abs{Q_{0,n}}$, $r\rightarrow r_n$, $A\rightarrow N$,
and $\beta\rightarrow \beta_n$.

To relate the deformation to the laboratory-frame (spectroscopic) observables,
we must return to the assumption of axially symmetric rotation.  Then quantity
$\abs{Q_{0,p}}$ appearing in~(\ref{eqn:Q0-r-beta-proton}) is equal, to within
sign, to the electromagnetic intrinsic quadrupole moment $Q_0$ appearing
in~(\ref{eqn:Q0-as-me}).\footnote{The quantities $\beta$ and $\abs{Q_0}$ are
interpreted, in~(\ref{eqn:beta-as-norm}) and~(\ref{eqn:Q0-as-norm}),
respectively, as the unsigned norms of spherical tensors.  However, the
intrinsic quadrupole moment $Q_0$ of~(\ref{eqn:Q0-as-me}) is a signed matrix
element, with sign related to the prolate or oblate nature of the deformation
described by the quadrupole tensor, and thus to the Bohr deformation parameter
$\gamma$~\cite{bohr1952:vibcoupling}.  A positive value of $Q_0$ indicates prolate
deformation ($\gamma=0$), and a negative value oblate deformation
($\gamma=\pi/3$).}  It is therefore simply related to the rotational $E2$
observables via Clebsch-Gordan cofficients (more general relations can be
deduced for triaxial~\cite{meyertervehn1975:triax-odda} or
$\grpsu{3}$~\cite{castanos1988:su3-shape} rotation).

We may connect the proton deformation $\beta_p$ directly with the dimensionless
ratio $Q/r_p^2$, for the spectroscopic quadrupole moment $Q(J)$, by
taking~(\ref{eqn:Q0-r-beta-proton}) in combination with the rotational relation (footnote~\ref{fn:rot-reln}), to obtain
\begin{equation}
  \label{eqn:Q-norm-r-beta-proton}
  \frac{\abs{Q(J)}}{Zr_p^2}
  =
  \Bigl(\frac{5}{\pi}\Bigr)^{1/2}
  \frac{3K^2-J(J+1)}{(J+1)(2J+3)}
  \beta_p.
\end{equation}
Thus, \textit{e.g.},
for a $K=3/2$ band head, 
\begin{math}
\beta_p=\sqrt{5\pi}\abs{Q(3/2)}/(Zr_p^2);
\end{math}
for the $J=3/2$ member of a $K=1/2$ band,
\begin{math}
\beta_p=\sqrt{5\pi}\abs{Q(3/2)}/(Zr_p^2);
\end{math}
or, for the $J=2$ member of a $K=0$ band,
\begin{math}
\beta_p=[7\sqrt{\pi}/(2\sqrt{5})] \abs{Q(2)}/(Zr_p^2).
\end{math}

Analogous relations may be derived for $B(E2)/(e^2r_p^4)$, which, again
from~(\ref{eqn:Q0-r-beta-proton}), is related to $\beta_p$, in the limit of
axially symmetric rotation, by
\begin{equation}
  \label{eqn:be2-norm-rp-beta-proton}
  \frac{B(E2;J_i\rightarrow J_f)}{Z^2e^2r_p^4}=\Bigl(\frac{5}{4\pi}\Bigr)^2
  \tcg{J_i}{K}{2}{0}{J_f}{K}^2 \beta_p^2.
\end{equation}
Specifically, for the
$0\rightarrow 2$ (upward) transition strength within a $K=0$ band, which is commonly
taken~\cite{raman2001:systematics,pritychenko2016:e2-systematics} as a
phenomenological measure of deformation,
we have
\begin{equation}
  \label{eqn:beta-be2-norm-r-proton-k0}
  \beta_p=\frac{4\pi}{5} \frac{B(E2;0\rightarrow2)^{1/2}}{Zer_p^2}.
\end{equation}
Note here that $B(E2;0\rightarrow2)/B(E2;2\rightarrow0)=5$.

By comparison, the traditional expression
\begin{equation}
  \label{eqn:beta-be2-norm-r-proton-k0-traditional}
  \beta = \frac{4\pi}{3}\frac{B(E2;0\rightarrow2)^{1/2}}{Zer_0^2A^{2/3}},
\end{equation}
used in Refs.~\cite{raman2001:systematics,pritychenko2016:e2-systematics}, is
obtained from~(\ref{eqn:beta-be2-norm-r-proton-k0}) if, instead of using the
actual nuclear radius, one adopts the global phenomenological estimate
$r=(3/5)^{1/2} r_0 A^{1/3}$, with $r_0=1.2\,\fm$, for the nuclear r.m.s.\ radius.
This global estimate does not distinguish between proton, neutron, or charge
radii.  It thus effectively assumes that charge is homogeneously distributed
within the nuclear matter (an asumption which is to be regarded with extreme
caution for light nuclei) and ignores local structural features.
 \begin{figure*}
\centering
\includegraphics[width=\ifproofpre{0.95}{0.95}\hsize]{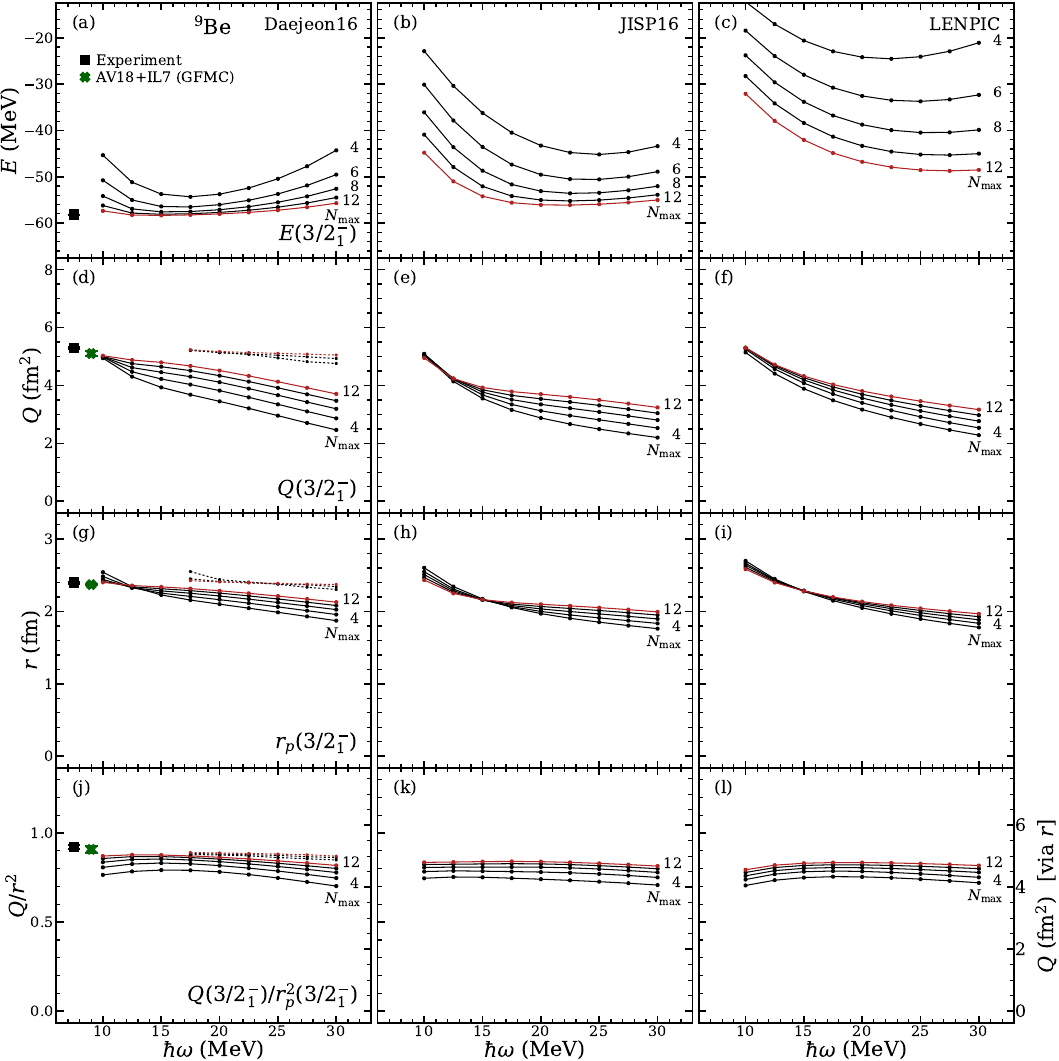}
\caption{Calculated ground state observables for $\isotope[9]{Be}$:
  $E(3/2^-_1)$~(top), $Q(3/2^-_1)$~(upper middle), $r_p(3/2^-_1)$~(lower
  middle), and the dimensionless ratio $Q/r_p^2$ constructed from the preceding
  two observables~(bottom).  Results are shown for the Daejeon16 (left), JISP16
  (center), and LENPIC (right) interactions. Calculated values are shown as
  functions of the basis parameter $\hw$, for successive even values of $\Nmax$,
  from $\Nmax=4$ to $12$ (as labeled).  When calibrated to the experimentally
  deduced value for $r_p$, the ratio provides a prediction for the absolute $Q$
  (scale at right).  Exponential extrapolations (small circles, dotted lines)
  are provided, for the Daejeon16 results only ($\hw\geq17.5\,\MeV$).  For
  comparison, experimental
  values~\cite{stone2016:e2-moments,angeli2013:charge-radii,npa2004:008-010}
  (squares) and GFMC AV18+IL7 predictions~\cite{pastore2013:qmc-em-alt9}
  (crosses) are also shown.  }
\label{fig:q-norm-rp-scan-9be}
\end{figure*}

\begin{figure*}
\centering
\includegraphics[width=\ifproofpre{0.85}{0.85}\hsize]{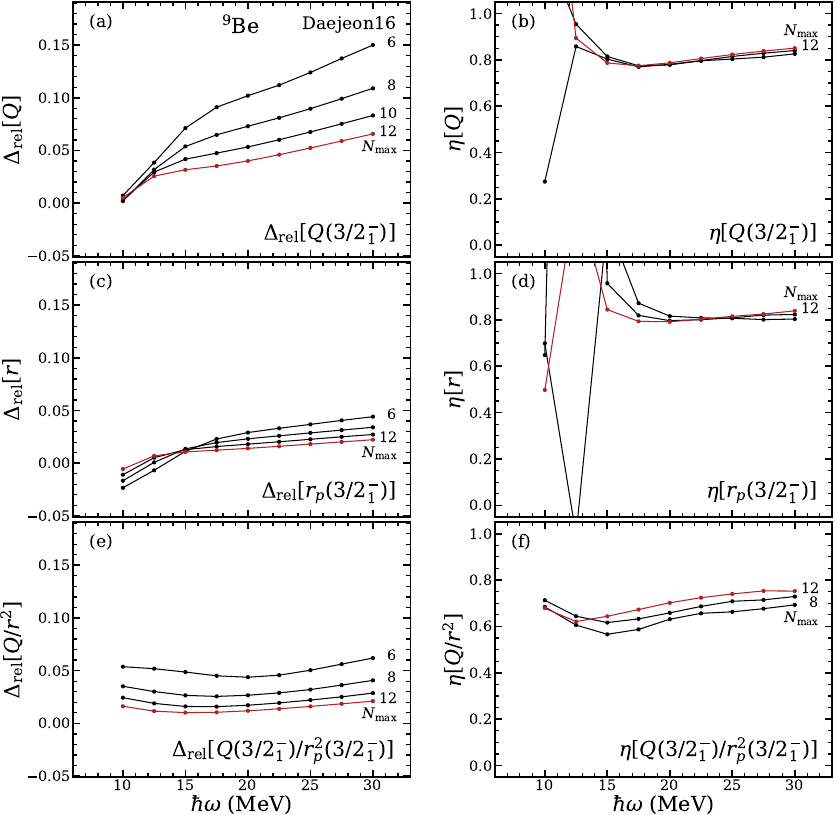}
\caption{Diagnostics of convergence for ground state observables for
  $\isotope[9]{Be}$: the relative difference $\Deltarel$~(left) and ratio of
  successive differences $\eta$~(right), for $Q(3/2^-_1)$~(top),
  $r_p(3/2^-_1)$~(middle), and the dimensionless ratio $Q/r_p^2$~(bottom).
  Calculated values, for the Daejeon16 interaction, are shown as functions of
  the basis parameter $\hw$, for successive even values of $\Nmax$, from
  $\Nmax=6$ or $8$ (as appropriate, given observables calculated starting with
  $\Nmax=4$) to $12$ (as labeled).  }
\label{fig:q-norm-rp-scan-9be-convergence-diagnostics}
\end{figure*}

\section{Convergence illustration: $\isotope[9]{Be}$}
\label{sec:results-moments:9be}

As an illustrative example, let us first take the $3/2^-$ ground state of
$\isotope[9]{Be}$.  The ground state charge radius and quadrupole moment are
both known experimentally (Fig.~\ref{fig:nuclear-chart}).

The calculated ground state observables are shown in
Fig.~\ref{fig:q-norm-rp-scan-9be}.  Calculations are carried out using the NCCI
code MFDn~\cite{maris2010:ncsm-mfdn-iccs10,*shao2018:ncci-preconditioned}.  The
basis for an NCCI calculation must be restricted to finite size, by truncation
to configurations with some maximum number $\Nmax$ of oscillator excitations
(taken relative to the lowest Pauli-allowed filling of oscillator
shells)~\cite{barrett2013:ncsm}.  Results converge towards the true result (that
is, as it would be obtained from solution of the many-body problem in the full,
untruncated many-body space) with increasing $\Nmax$.  Each curve in
Fig.~\ref{fig:q-norm-rp-scan-9be} represents the results of calculations sharing
the same $\Nmax$, but with varying choices of the oscillator parameter $\hw$,
which determines the oscillator length
$b\propto(\hw)^{-1/2}$~\cite{suhonen2007:nucleons-nucleus}, for the basis
functions.  (Results are shown for $4\leq\Nmax\leq12$ and $10\,\MeV\leq
\hw\leq30\,\MeV$.) In general, an approach to convergence is signaled by
calculated values which no longer change with increasing $\Nmax$ (compression of
successive curves) and are locally insensitive to the choice of oscillator
parameter (flatness with respect to $\hw$).  Computational resources limit the
$\Nmax$ for which calculations are feasible and thus the precision which can be
obtained.

For purposes of comparison, we take NCCI calculations based on three different
internucleon interactions (left to right in Fig.~\ref{fig:q-norm-rp-scan-9be}):
Daejeon16~\cite{shirokov2016:nn-daejeon16},
JISP16~\cite{shirokov2007:nn-jisp16}, and a LENPIC chiral effective field theory
($\chi$EFT)
interaction~\cite{epelbaum2015:lenpic-n4lo-scs,epelbaum2015:lenpic-n3lo-scs}.
Of these interactions, Daejeon16 is the ``softest'', providing the most
favorable convergence properties.  It was obtained as the two-body part of an
\nthreelo{} $\chi$EFT interaction~\cite{entem2003:chiral-nn-potl}, but was then
softened via a similarity renormalization group (SRG)
transformation~\cite{bogner2007:srg-nucleon} so as to provide comparatively
rapid convergence (and subsequently adjusted via a phase-shift equivalent
transformation to better describe nuclei with $A\leq16$ without requiring the
introduction of a three-body contribution).  In contrast, the ``hardest'' is the
LENPIC interaction, which, for purposes of illustration, is taken here as a bare
interaction, with no SRG softening (specifically, we use the two-body
\ntwolo{} interaction with $R=1\,\fm$ semi-local coordinate-space
regulator~\cite{epelbaum2015:lenpic-n4lo-scs,epelbaum2015:lenpic-n3lo-scs}).

The difference in hardness of these interactions is readily apparent in the
calculated ground state energies, for which the convergence behavior is shown in
Fig.~\ref{fig:q-norm-rp-scan-9be} (top).  In each case, the calculated ground
state energy for a given fixed $\Nmax$ has a variational minimum at some $\hw$,
ranging from $\hw\approx15\,\MeV$ for Daejeon16
[Fig.~\ref{fig:q-norm-rp-scan-9be}(a)] to $\hw\lesssim30\,\MeV$ for LENPIC
[Fig.~\ref{fig:q-norm-rp-scan-9be}(c)].  The results for Daejeon16 display a
robust approach to convergence, at the level of $\lesssim0.1\,\MeV$ (for $\hw$
near the variational minimum), while the results for LENPIC
[Fig.~\ref{fig:q-norm-rp-scan-9be}(c)] still differ by $\approx5\,\MeV$ for the
highest two $\Nmax$ values shown.

Let us focus first on the results obtained with the Daejeon16 interaction, for
which the calculated quadrupole moment is shown in
Fig.~\ref{fig:q-norm-rp-scan-9be}(d).  These results show tendencies towards
convergence.  We can see a flattening (shouldering) of the curves, in the lower
portion of the $\hw$ range shown, and the spacing between curves for successive
$\Nmax$ decreases modestly.  It would not be unreasonable to estimate the
calculations to be converging towards something close to the experimental value
(square).  However, the convergence is not sufficiently developed for us to read
off a concrete estimate of the true result for the full,
untruncated space.  Although the curves appear to approximately cross at a
single point at the low end of the $\hw$ range shown, and it has been
suggested~\cite{bogner2008:ncsm-converg-2N} that such a crossing point may be
taken to provide a heuristic estimate for the true, converged value, such
estimates are not necessarily found to be robust~\cite{caprio2014:cshalo}.

Taking now the calculated point-proton radius, shown in
Fig.~\ref{fig:q-norm-rp-scan-9be}(g), the $\hw$ dependence is superficially
less pronounced than for the quadrupole moment
[Fig.~\ref{fig:q-norm-rp-scan-9be}(d)].  However, recall that the radius goes as
the square root of the matrix element of an operator with $r^2$ radial
dependence, while the quadrupole moment is simply proportional to such a matrix
element, and higher powers amplify relative changes (in this case, by a factor
of $2$).  For both observables, the values must rise towards infinity at
sufficiently low $\hw$ and must fall towards zero at sufficiently large $\hw$,
in a finite oscillator basis, since the oscillator functions only provide
support at a radial distance scale in the vicinity of the oscillator length
[$b\propto(\hw)^{-1/2}$].  The curves representing the radius for different
$\Nmax$ have a crossing point somewhat higher in $\hw$ (by $\approx2.5\,\MeV$)
than for the quadrupole moment.  This makes it clear that the calculated $Q$ and
$r_p^2$ cannot be strictly correlated, across different $\Nmax$ and $\hw$, as
they would have to be in order to yield a truly constant value for the dimensionless
ratio $Q/r_p^2$.

Nonetheless, taking the dimensionless ratio $Q/r_p^2$, as shown in
Fig.~\ref{fig:q-norm-rp-scan-9be}(j), serves to eliminate much of the $\hw$
dependence found in the calculated quadrupole moment.  Moreover, it would appear
to improve convergence with respect to $\Nmax$.  It may be noted that the ratio
$Q/r^2$ is monotonically increasing with $\Nmax$, with no crossings of curves
for successive $\Nmax$, over the $\hw$ range shown.

Such observations may be made more quantitative by considering the
\textit{differences} in the calculated values, for successive $\Nmax$, yielding
diagnostics of the convergence such as those shown (for these Daejeon16 results)
in Fig.~\ref{fig:q-norm-rp-scan-9be-convergence-diagnostics}.  For an observable
$X(\Nmax)$, at fixed $\hw$, we consider the relative difference (or logarithmic
difference) $\Deltarel[X]$
[Fig.~\ref{fig:q-norm-rp-scan-9be-convergence-diagnostics} (left)], which we
define by
\begin{equation}
  \label{eqn:relative-difference}
  \Deltarel[X](\Nmax)=\frac{X(\Nmax)-X(\Nmax-2)}{\tfrac12[X(\Nmax)+X(\Nmax-2)]}.
\end{equation}
We also consider the ratio of successive differences~\cite{fasano2022:natorb},
\begin{equation}
  \label{eqn:difference-ratio}
  \eta[X](\Nmax)\equiv \frac{X(\Nmax)-X(\Nmax-2)}{X(\Nmax-2)-X(\Nmax-4)},
\end{equation}
which measures how rapidly
the step size, between values for the observable for successive $\Nmax$, decreases with
increasing $\Nmax$.  That is, more concisely,
\begin{equation}
  \label{eqn:difference-ratio-from-delta}
  \eta[X](\Nmax) = \frac{\Delta[X](\Nmax)}{\Delta[X](\Nmax-2)},
\end{equation}
in terms of differences
\begin{equation}
  \label{eqn:difference}
  \Delta[X](\Nmax) = X(\Nmax) - X(\Nmax-2).
\end{equation}

If one assumes exponential
convergence with respect to $\Nmax$~\cite{forssen2008:ncsm-sequences,bogner2008:ncsm-converg-2N,maris2009:ncfc}, of the form
\begin{equation}
  \label{eqn:exp-convergence}
  X(\Nmax)=X_{\infty}+a\exp(-c\Nmax),
\end{equation}
then the limiting value $X_{\infty}$ is approached (as $\Nmax\rightarrow\infty$)
in steps the sizes of which form a geometric progression, and the ratio of
successive differences is simply a constant $\eta[X]=e^{-2c}$, independent of
$\Nmax$.  Conversely, constant $\eta$ indicates exponential convergence.  Some
observations regarding this geometric progression, relevant to carrying out a
three-point exponential extrapolation, may be found in
Appendix~\ref{sec:app-geometric}.

For the quadrupole moment itself, the relative differences $\Deltarel[Q]$
[Fig.~\ref{fig:q-norm-rp-scan-9be-convergence-diagnostics}(a)] essentially
vanish ($\lesssim1\%$) at the low end of the $\hw$ range shown, where the curves
for $Q$ [Fig.~\ref{fig:q-norm-rp-scan-9be}(d)] are crossing.  Taking
$\hw=20\,\MeV$ as a more illustrative point for comparison, the relative
differences decrease from $\approx 10\%$ at $\Nmax=6$ to $\approx 4\%$ at
$\Nmax=12$.  For the radius, the relative differences $\Deltarel[r_p]$
[Fig.~\ref{fig:q-norm-rp-scan-9be-convergence-diagnostics}(c)] range from
$\approx 3\%$ at $\Nmax=6$ to $\approx 1.4\%$ at $\Nmax=12$, again for
$\hw=20\,\MeV$, reflecting both the different $\hw$ at which the curves cross
[Fig.~\ref{fig:q-norm-rp-scan-9be}(g)] and a smaller overall scale of
differences.  At least in the region of larger $\hw$, to the right of any
crossing points, the ratios of successive differences, $\eta[Q]$
[Fig.~\ref{fig:q-norm-rp-scan-9be-convergence-diagnostics}(b)] and $\eta[r_p]$
[Fig.~\ref{fig:q-norm-rp-scan-9be-convergence-diagnostics}(d)], demonstrate that
the convergence is approximately exponential for both the quadrupole moment and
radius, with a ratio $\eta\approx 0.8$ for successive steps, or an $\approx20\%$
reduction in size for successive steps.  This implies that the distance
remaining to the limiting value is $\approx 4$ times the size of the last step
taken (see Appendix~\ref{sec:app-geometric}).  This ratio, and thus the decay
constant $c$, is only weakly dependent on $\hw$, increasing to $\approx0.85$ by
$\hw=30\,\MeV$.

Then, for the ratio the relative differences $\Deltarel[Q/r^2]$
[Fig.~\ref{fig:q-norm-rp-scan-9be-convergence-diagnostics}(e)] are relatively
independent of $\hw$, decreasing from $\approx 4\%$ at $\Nmax=6$ to $\approx
1.2\%$ at $\Nmax=12$, much smaller than for the quadrupole moment itself over
most of the $\hw$ range
[Fig.~\ref{fig:q-norm-rp-scan-9be-convergence-diagnostics}(a)].  The ratios
$\eta[Q/r^2]$ of successive differences
[Fig.~\ref{fig:q-norm-rp-scan-9be-convergence-diagnostics}(f)] indicate that the
convergence pattern for $Q/r_p^2$ less closely matches exponential convergence
in $\Nmax$, as $\eta$ varies significantly with $\Nmax$, increasing from
$\approx 0.6$ to $\approx0.7$ over the three steps in $\Nmax$ shown. Indeed, the
ratio of two exponentially converging quantities is not itself, in general,
expected to be exponentially converging, although it may be approximately so in
limiting cases.

Returning to the other interactions in Fig.~\ref{fig:q-norm-rp-scan-9be}, any
indication of shouldering in the calculated quadrupole moment is less visible
for JISP16 [Fig.~\ref{fig:q-norm-rp-scan-9be}(e)] and especially for the bare
LENPIC interaction [Fig.~\ref{fig:q-norm-rp-scan-9be}(f)].  The successive
differences in the calculated quadrupole moment do decrease in size (in fact,
the convergence for JISP16 is again remarkably exponential in $\Nmax$, with
$\eta\approx0.8$).  But, again, we would be hard put to read off a concrete
estimate of the true result.  Nonetheless, even for these harder interactions,
taking the dimensionless ratio $Q/r_p^2$
[Fig.~\ref{fig:q-norm-rp-scan-9be}(k,l)] once again serves to eliminate much of
the $\hw$ dependence found in the calculated quadrupole moment.

Moreover, for the three choices of interaction considered here, the calculated
ratios $Q/r_p^2$ [Fig.~\ref{fig:q-norm-rp-scan-9be} (bottom)] are largely
indistinguishable.  In an axially symmetric rotational description, we may
interpret this as a robust prediction for the deformation
(Sec.~\ref{sec:background:deformation}), independent of choice of interaction.
The calculated ratios, for all three interactions, lie below the experimental
ratio (square) but increase towards it from below with increasing $\Nmax$, such
that there is no clear discrepancy between the predictions and experiments.  The
experimental ratio $Q/r_p^2\approx0.92$ yields, via~(\ref{eqn:Q-norm-r-beta-proton}),
a deformation $\beta_p\approx0.91$.

While our goal in the present work is not to carry out detailed studies of
prospective basis extrapolation schemes (\textit{e.g.},
Ref.~\cite{odell2016:ir-extrap-quadrupole}), the above observations
qualitatively consistency with exponential convergence suggest attempting at
least a basic three-point exponential extrapolation with respect to $\Nmax$
(Appendix~\ref{sec:app-geometric}).  The results of such an extrapolation are
shown (small circles, dotted lines), for the $\isotope[9]{Be}$ results obtained
with the Daejeon16 interaction, for $Q$ [Fig.~\ref{fig:q-norm-rp-scan-9be}(d)],
$r_p^2$ [Fig.~\ref{fig:q-norm-rp-scan-9be}(g)], and $Q/r_p^2$
[Fig.~\ref{fig:q-norm-rp-scan-9be}(j)].  (From the calculations considered here,
which have $\Nmax\geq4$, a three-point extrapolation becomes possible for
$\Nmax\geq8$.)

While we cannot compare the extrapolated values to a true, converged result, we
may note that the extrapolations are remarkably stable with respect to both
$\Nmax$ and $\hw$, in the region ($\hw\gtrsim17.5\,\MeV$) well above the
location of any crossing of the curves, at least for the underlying observables
$Q$ [Fig.~\ref{fig:q-norm-rp-scan-9be}(d)] and $r_p^2$
[Fig.~\ref{fig:q-norm-rp-scan-9be}(g)].  Such an extrapolation is less obviously
useful for the ratio [Fig.~\ref{fig:q-norm-rp-scan-9be}(j)], seen above
[Fig.~\ref{fig:q-norm-rp-scan-9be-convergence-diagnostics}(f)] to be more
rapidly convergent but less exponential in its convergence.

Similar extrapolations for the results obtained with the JISP interaction are
more erratic, and those obtained with the (bare) LENPIC interactions are
essentially unusable.  (These extrapolations are not shown in
Fig.~\ref{fig:q-norm-rp-scan-9be}, to avoid obscuring the curves of principal
interest, but are available in the Supplemental
Material~\cite{supplemental-material}.)  This difference in success of the
extrapolation might be taken to reflect the less converged starting point for
extrapolation provided by these harder interactions.

The ground state quadrupole moment is known for $\isotope[9]{Be}$, so that here
our goal is to provide an experimental test of the predicted ratio, rather than
an estimate for an unknown quadrupole moment from a known radius (or
\textit{vice versa}).  Nonetheless, for illustration, normalizing to the
experimental proton radius gives the scale for $Q$ shown at right in
Fig.~\ref{fig:q-norm-rp-scan-9be}~(bottom).

A point of theoretical comparison is provided by the Green's function Monte
Carlo (GFMC)~\cite{carlson2015:qmc-nuclear} approach, which also yields
predictions for $E2$ and radius observables for lower $p$-shell nuclei.  The
predicted $Q$ and $r_p$ for $\isotope[9]{Be}$, from GFMC
calculations~\cite{pastore2013:qmc-em-alt9} with the Argonne $v_{18}$ (AV18)
two-nucleon~\cite{wiringa1995:nn-av18} and Illinois-7 (IL7)
three-nucleon~\cite{pieper2008:3n-il7-fm50} potentials, are shown (crosses) in
Fig.~\ref{fig:q-norm-rp-scan-9be}, along with the deduced ratio $Q/r_p^2$
[Fig.~\ref{fig:q-norm-rp-scan-9be}~(bottom)].  In the GFMC results, the dominant
uncertainties are statistical in nature, and errors would not \textit{a priori}
be expected to cancel in the ratio (although some correlation in errors is
possible, especially if the GFMC calculations for different observables are
carried out using the same set of Monte Carlo samples).  The ratio is only taken
here for purposes of comparison.  The GFMC predictions for both observables
individually lie just below experiment, while the ratio is consistent with
experiment to within uncertainties.  The NCCI results therefore also appear to
be converging towards approximate consistency with the GFMC result for the
ratio.

 \begin{figure*}
\centering
\includegraphics[width=\ifproofpre{0.70}{0.75}\hsize]{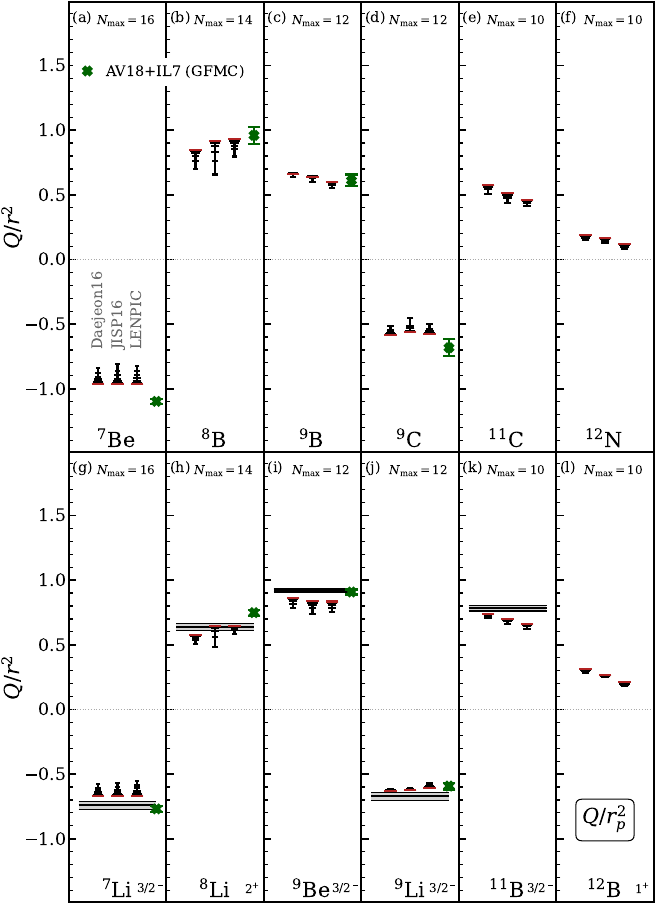}
\caption{Calculated ratios $Q/r_p^2$, for the ground states of
  proton-rich~(top) and neutron-rich~(bottom) nuclides in the $p$ shell,
  obtained with the Daejeon16, JISP16, and LENPIC interactions (from left to
  right, within each panel).  Calculated values are shown at fixed
  $\hw=20\,\MeV$ and varying $\Nmax$ (increasing tick size), from $\Nmax=4$ to
  the maximum value indicated (at top).  For comparison, the experimental
  ratios~\cite{stone2016:e2-moments,angeli2013:charge-radii} are shown
  (horizontal line and error band, where the signs of some quadrupole moments
  are experimentally undetermined), as are the GFMC AV18+IL7
  predictions~\cite{pastore2013:qmc-em-alt9,pieper:cited} (crosses). }
\label{fig:q-norm-rp-ratio-teardrop-mirror}
\end{figure*}

\begin{figure}
\centering
\includegraphics[width=\ifproofpre{0.8}{0.6}\hsize]{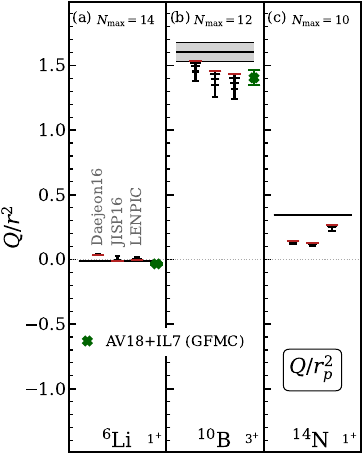}
\caption{Calculated ratios $Q/r_p^2$, for the ground states of $N=Z$
  nuclides in the $p$ shell, obtained with the Daejeon16, JISP16, and LENPIC
  interactions (from left to right, within each panel).  Calculated values are shown at fixed
  $\hw=20\,\MeV$ and varying $\Nmax$ (increasing tick size), from $\Nmax=4$ to
  the maximum value indicated (at top).  For comparison, the experimental
  ratios~\cite{stone2016:e2-moments,angeli2013:charge-radii} are shown
  (horizontal line and error band, where the signs of some quadrupole moments
  are experimentally undetermined), as are the GFMC AV18+IL7
  predictions~\cite{pastore2013:qmc-em-alt9} (crosses).}
\label{fig:q-norm-rp-ratio-teardrop-neqz}
\end{figure}

\begin{figure*}
\centering
\includegraphics[width=\ifproofpre{0.70}{0.75}\hsize]{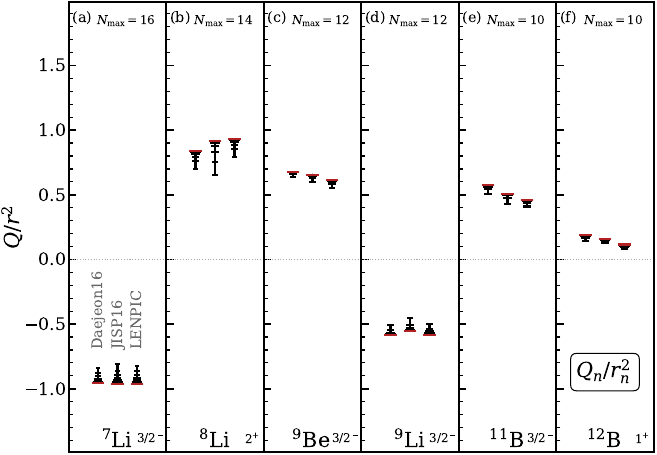}
\caption{Calculated ratios $Q_n/r_n^2$, involving neutron observables,
  for the ground states of neutron-rich nuclides in the $p$ shell, obtained with
  the Daejeon16, JISP16, and LENPIC interactions (from left to right, within
  each panel).  Calculated values are shown at fixed $\hw=20\,\MeV$ and varying
  $\Nmax$ (increasing tick size), from $\Nmax=4$ to the maximum value indicated
  (at top).  }
\label{fig:q-norm-rp-ratio-teardrop-mirror-nrich-n}
\end{figure*}

\begin{figure*}
  \centering
\includegraphics[width=\ifproofpre{0.75}{1.0}\hsize]{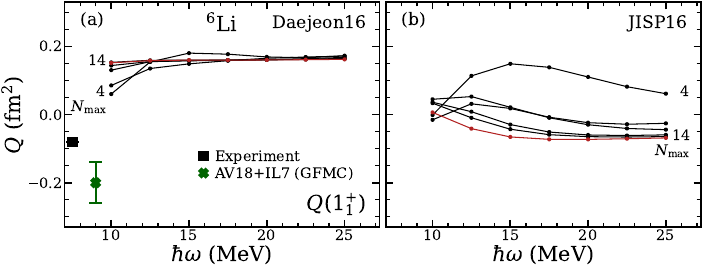}
\caption{Calculated ground state quadrupole moment 
  $Q(1^+_1)$ for $\isotope[6]{Li}$.  Results
  are shown for the (a)~Daejeon16 and (b)~JISP16
  interactions. Calculated values are shown as functions of the basis parameter
  $\hw$, for successive even values of $\Nmax$, from $\Nmax=4$ to $14$ (as
  labeled).  For
  comparison, the experimental
  value~\cite{stone2016:e2-moments} (square) and GFMC
  AV18+IL7 prediction~\cite{pastore2013:qmc-em-alt9} (cross) are also shown.
  }
\label{fig:q-scan-6li}
\end{figure*}

\begin{figure*}
\centering
\includegraphics[width=\ifproofpre{0.9}{0.9}\hsize]{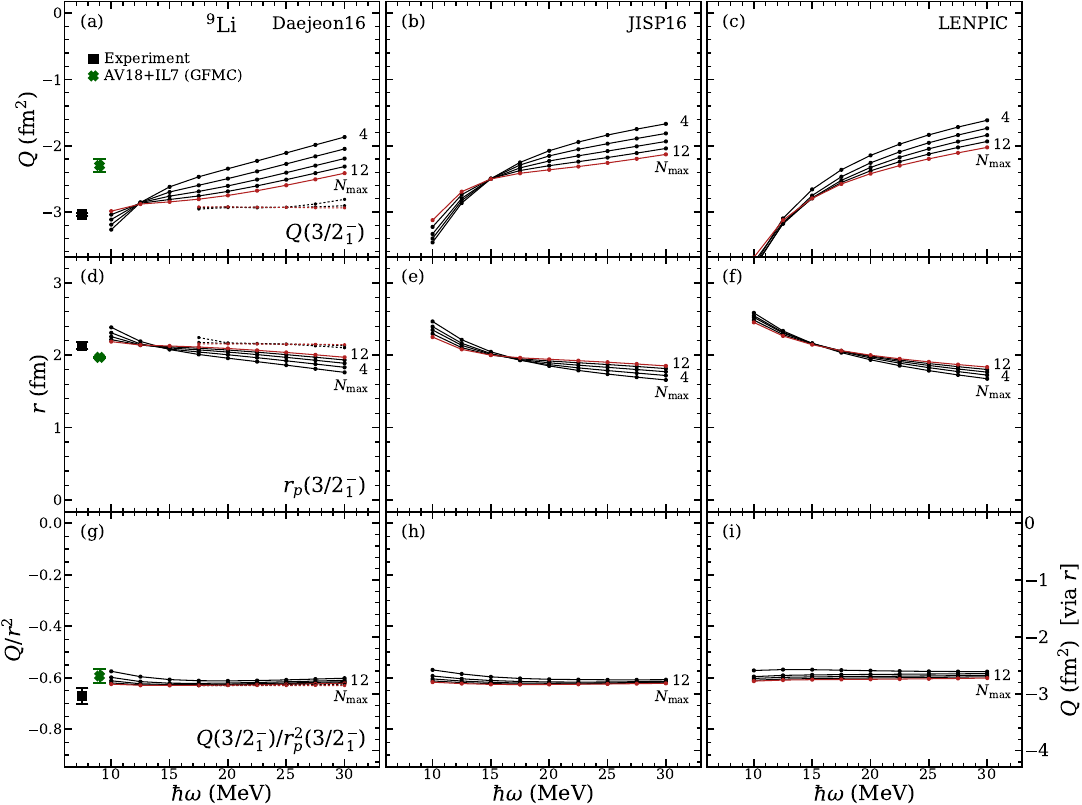}
\caption{Calculated ground state observables for $\isotope[9]{Li}$:
  $Q(3/2^-_1)$~(top), $r_p(3/2^-_1)$~(middle), and the dimensionless ratio
  $Q/r_p^2$~(bottom).  Results are shown for the (left)~Daejeon16,
  (center)~JISP16, and (right)~LENPIC interactions. Calculated values are shown
  as functions of the basis parameter $\hw$, for successive even values of
  $\Nmax$, from $\Nmax=4$ to $12$ (as labeled).  When calibrated to the
  experimentally deduced value for $r_p$, the ratio provides a prediction for
  the absolute $Q$ (scale at right).  Exponential extrapolations (small circles,
  dotted lines) are provided, for the Daejeon16 results only
  ($\hw\geq17.5\,\MeV$).  For comparison, experimental
  values~\cite{stone2016:e2-moments,angeli2013:charge-radii} (squares) and GFMC
  AV18+IL7 predictions~\cite{pastore2013:qmc-em-alt9} (crosses) are also shown.
}
\label{fig:q-norm-rp-scan-9li}
\end{figure*}

\begin{figure*}
\centering
\includegraphics[width=\ifproofpre{0.9}{0.9}\hsize]{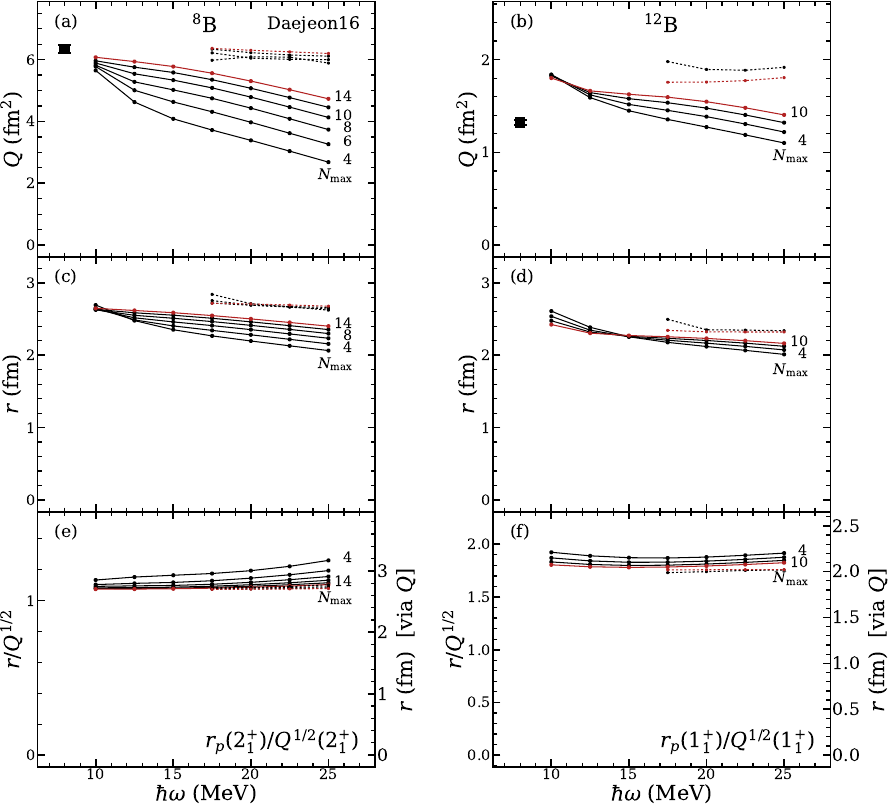}
\caption{Calculated observables for the $2^+$ ground state of $\isotope[8]{B}$~(left) and the $1^+$ ground state of $\isotope[12]{B}$~(right):
  $Q$~(top), $r_p$~(middle), and the dimensionless ratio $r_p/Q^{1/2}$~(bottom).  Results are shown for the Daejeon16 interaction.
  Calculated values are shown as functions of the basis parameter $\hw$, for
  successive even values of $\Nmax$, from $\Nmax=4$ to the maximum value calculated (as labeled).  For
  comparison, the experimental quadrupole moment~\cite{stone2016:e2-moments} is
  shown (horizontal line and error band).  When calibrated to the experimentally
  deduced value for $Q$, the ratio provides a prediction for the absolute $r_p$
  (scale at right).  Exponential extrapolations (small circles, dotted lines)
  are provided ($\hw\geq17.5\,\MeV$).
}
\label{fig:q-norm-rp-scan-8b-12b}
\end{figure*}

\section{Quadrupole moments of $p$-shell nuclei}
\label{sec:results-moments:survey}

For a concise overview of the \textit{ab initio} results for $Q/r_p^2$ across
the $p$ shell, in Figs.~\ref{fig:q-norm-rp-ratio-teardrop-mirror}
and~\ref{fig:q-norm-rp-ratio-teardrop-neqz} we restrict our attention to a
single, fixed value for the oscillator parameter $\hw$, and examine the $\Nmax$
dependence of the calculated results.  (Here, we take $\hw=20\,\MeV$ as a
representative value, near the variational energy minima obtained for the
Daejeon16 and JISP16 interactions [Fig.~\ref{fig:q-norm-rp-scan-9be} (top)].)
Results for mirror nuclide pairs are shown in
Fig.~\ref{fig:q-norm-rp-ratio-teardrop-mirror} (with proton-rich nuclei at top,
and neutron-rich nuclei at bottom), while results for $N=Z$ nuclei are shown in
Fig.~\ref{fig:q-norm-rp-ratio-teardrop-neqz}.\footnote{The convergence of the
quadrupole moment itself may be seen in the Supplemental
Material~\cite{supplemental-material}.  See also Fig.~4 of
Ref.~\cite{caprio2021:emratio}.}  In addition to the ground states of
particle-bound $p$-shell nuclides (with $J\geq1$), we also include, to complete
the set of mirror nuclides, the narrow ground state resonance of
$\isotope[9]{B}$~[Fig.~\ref{fig:q-norm-rp-ratio-teardrop-mirror}(c)].

Thus, for $\isotope[9]{Be}$, the
results in Fig.~\ref{fig:q-norm-rp-ratio-teardrop-mirror}(i) represent a
vertical slice through the curves shown in
Fig.~\ref{fig:q-norm-rp-scan-9be}~(bottom).  Comprehensive plots of the
calculated (and extrapolated) observables and ratios, as functions of
both $\Nmax$ and $\hw$, for both proton and neutron observables, are provided in
the Supplemental Material~\cite{supplemental-material}, along with numerical
tabulations of the calculated quantities.

Both the quadrupole moment and radius are simultaneously known, yielding an
experimental value for the dimensionless ratio, for all the neutron-rich nuclei
except $\isotope[12]{B}$
[Fig.~\ref{fig:q-norm-rp-ratio-teardrop-mirror}~(bottom)], and for the $N=Z$
nuclei (Fig.~\ref{fig:q-norm-rp-ratio-teardrop-neqz}), but not for any of the
proton-rich nuclei [Fig.~\ref{fig:q-norm-rp-ratio-teardrop-mirror}~(top)].  The
experimental results for the ratio are shown in
Figs.~\ref{fig:q-norm-rp-ratio-teardrop-mirror}
and~\ref{fig:q-norm-rp-ratio-teardrop-neqz} (horizontal lines, with error bands)
where available. (The experimental quadrupole moments in these figures are from
Ref.~\cite{stone2016:e2-moments}, except that the value for $\isotope[12]{N}$ is
corrected as noted in Table~I of Ref.~\cite{caprio2021:emratio}.  The
experimental point-proton radii are deduced, as described in
Sec.~\ref{sec:background:rc}, from the evaluated charge radii of
Ref.~\cite{angeli2013:charge-radii}.)  The GFMC AV18+IL7
predictions~\cite{pastore2013:qmc-em-alt9,pieper:cited}, are also shown
(crosses), for $A\leq 10$.

We may note some overall features of the calculated ratios $Q/r_p^2$
(Figs.~\ref{fig:q-norm-rp-ratio-teardrop-mirror}
and~\ref{fig:q-norm-rp-ratio-teardrop-neqz}):

(1)~The magnitude of the calculated ratio (that is, regardless of sign)
typically converges from below, that is, monotonically increases in $\Nmax$,
except in the anomalous case of $\isotope[6]{Li}$
[Fig.~\ref{fig:q-norm-rp-ratio-teardrop-neqz}(a)] (discussed below).

(2)~Convergence of the calculated ratio tends to be slower for the proton-rich
nuclei [Fig.~\ref{fig:q-norm-rp-ratio-teardrop-mirror} (top)], than for their
mirror neutron-rich nuclei [Fig.~\ref{fig:q-norm-rp-ratio-teardrop-mirror}
  (bottom)], with the exception of the $\isotope[9]{B}$/$\isotope[9]{Be}$ mirror
pair [Fig.~\ref{fig:q-norm-rp-ratio-teardrop-mirror}(c,i)].  In particular, this
difference is seen in the greater spread among the calculated values (or longer
``tail'' in the plot), from lowest to highest $\Nmax$ shown.

(3)~The predictions for the ratio rarely differ, among the three interactions,
by more than $\approx0.1$, with the exception of $\isotope[14]{N}$
[Fig.~\ref{fig:q-norm-rp-ratio-teardrop-neqz}(c)].  However, strict claims
cannot be made, given incompletely converged results.

(4)~Where the ratio is experimentally known, the calculated ratios are in
general agreement with experiment, typically differing by $\lesssim0.1$ (or at
most $\approx0.2$, for $\isotope[14]{N}$
[Fig.~\ref{fig:q-norm-rp-ratio-teardrop-neqz}(c)]) from the experimental central
value.  If there is an overall tendency, it is that the calculations
underpredict the ratio (in magnitude) rather than overpredict.  (The
difference between the calculated and measured ratio is perhaps most striking
for $\isotope[7]{Li}$ [Fig.~\ref{fig:q-norm-rp-ratio-teardrop-mirror}(g)], not
for the scale of the disagreement, but since the NCCI results are so
consistent across interactions, with
differences among themselves much smaller than the difference from experiment.)
However, since
the calculated values increase (in magnitude) with $\Nmax$, this observation may
in part be an artifact of incomplete convergence.  Furthermore, any differences
from experiment should be revisited in the context of the corrections ommitted
in translating the charge radius to a point-proton radius
(Sec.~\ref{sec:background:rc}), as well as possible meson exchange (or $\chi$EFT) corrections
to the $E2$ operator used in calculating the quadrupole moment.

(5)~The calculated ratios are also in approximate agreement with those from the
GFMC AV18+IL7 calculations, again differing by $\lesssim0.1$, allowing for the
statistical uncertainties (shown) in the GFMC results.  A notable discrepancy is
found for the $A=7$ nuclei [Fig.~\ref{fig:q-norm-rp-ratio-teardrop-mirror}(a,g)]
where the NCCI calculations for different interactions are in close agreement
with each other (as just noted above for $\isotope[7]{Li}$), but the GFMC result
is larger (in magnitude) by $\approx0.1$.  A similar discrepancy is found for
$\isotope[8]{Li}$ [Fig.~\ref{fig:q-norm-rp-ratio-teardrop-mirror}(h)], but not
the mirror nuclide $\isotope[8]{B}$
[Fig.~\ref{fig:q-norm-rp-ratio-teardrop-mirror}(b)].

Returning to the comparison across mirror nuclei, in the second item above,
differences in convergence rate between the neutron-rich nuclei
[Fig.~\ref{fig:q-norm-rp-ratio-teardrop-mirror} (bottom)] and their proton-rich
mirror nuclei [Fig.~\ref{fig:q-norm-rp-ratio-teardrop-mirror} (top)] need not
imply any differences in convergence rates for the wave functions \textit{per
  se}, for these two sets of nuclei.  Rather, it must be kept in mind that the
observables under consideration probe only the protons.  However, the protons
are the more tightly bound nucleonic species in the neutron-rich nuclei, but the
less tightly bound species in the proton-rich nuclei.  Spatially extended halo
orbitals or molecular orbitals (\textit{e.g.},
Ref.~\cite{seya1981:molecular-orbital,freer2007:cluster-structures}) for
neutrons are expected to be prevalent in neutron-rich nuclei, with mirror
symmetry implying that protons are to be found in corresponding spatially
extended configurations in proton-rich nuclei (\textit{e.g.},
Ref.~\cite{henninger2015:8b-fmd}).

Indeed, comparing the calculated ratio of
neutron observables $Q_n/r_n^2$ in the neutron-rich nuclei, as shown in
Fig.~\ref{fig:q-norm-rp-ratio-teardrop-mirror-nrich-n}, with the corresponding
ratio of proton observables for the proton-rich mirror nuclei
[Fig.~\ref{fig:q-norm-rp-ratio-teardrop-mirror} (top)], we see qualitative
features which are essentially identical.  Thus the main differences in
$Q/r_p^2$ across mirror nuclides
(Fig.~\ref{fig:q-norm-rp-ratio-teardrop-mirror}), both in value and in
convergence behavior, follow simply from the proton-neutron asymmetry of the
structure, while maintaining mirror symmetry.\footnote{Residual quantitative deviations
from mirror symmetry do arise, due to effects of the Coulomb interaction in the
Daejeon16 and JISP16 calculations, as well as additional strong-interaction components
in the LENPIC calculations.  The extent to which mirror symmetry is violated in
the calculated quadrupole moments is examined in detail in Sec.~IV of
Ref.~\cite{caprio2021:emratio}.}

The nuclide $\isotope[6]{Li}$ [Fig.~\ref{fig:q-norm-rp-ratio-teardrop-neqz}(a)]
warrants comment as an exceptional case.  The experimental quadrupole moment
[$Q(1^+)\approx-0.08\,\fm^2$] is an order of magnitude smaller than suggested by
the Weisskopf single-particle scale~\cite{weisskopf1951:estimate} of
$\approx1\,e\fm^2$ for $E2$ matrix elements in this mass region, and
$1\text{--}2$ orders of magnitude smaller than the other measured quadrupole
moments for $p$-shell nuclides~\cite{stone2016:e2-moments}.  The \textit{ab
  initio} calculations for the $\isotope[6]{Li}$ quadrupole moment have atypical
convergence patterns with $\Nmax$ and $\hw$, as shown in
Fig.~\ref{fig:q-scan-6li} for the Daejeon16 and JISP16 interactions.
(Convergence of the $\isotope[6]{Li}$ quadrupole moment for the JISP16
interaction was explored in detail in
Refs.~\cite{cockrell2012:li-ncfc,maris2013:ncsm-pshell}.)  The NCCI calculated
quadrupole moment starts positive at low $\Nmax$.  The Daejeon16 results
[Fig.~\ref{fig:q-scan-6li}(a)] then rapidly converge to a smaller, but still
positive, value of $Q(1^+)\approx+0.15\,\fm^2$, while the JISP16 results
[Fig.~\ref{fig:q-scan-6li}(b)] cross zero (as do the even less monotonic LENPIC
results, not shown).  Changes so large and erratic that they even affect the
sign are not the type of systematic truncation error which we might hope to
offset by simply dividing out the smooth convergence behavior of the squared
radius.  (The value of the quadrupole moment in $\isotope[6]{Li}$, much smaller
than the typical single-particle scale, may be understood as arising from
cancellations of short-range and long-range contributions in the radial
quadrupole density~\cite{cockrell2012:li-ncfc}, which may also underlie the
atypical convergence behavior.)  Doing so, as in
Fig.~\ref{fig:q-norm-rp-ratio-teardrop-neqz}(a), can have only an incidental
effect in ameliorating the irregularities.

Although the convergence for $\isotope[9]{Li}$, shown in
Fig.~\ref{fig:q-norm-rp-scan-9li}, is similar to that  for
$\isotope[9]{Be}$ (Fig.~\ref{fig:q-norm-rp-scan-9be}), further comment is warranted on the magnitude of the quadrupole moment.
We again find that taking the dimensionless ratio $Q/r_p^2$
[Fig.~\ref{fig:q-norm-rp-scan-9li} (bottom)] largely eliminates the $\hw$
dependence, while also improving compression of the results for successive
$\Nmax$.  For $\isotope[9]{Li}$, the predicted ratio $Q/r_p^2\approx-0.63$ is
again consistent across all three interactions (although somewhat less well
converged for LENPIC).  These values lie just below (in magnitude) the
uncertainties on the experimental ratio
$Q/r_p^2=-0.67(3)$~\cite{stone2016:e2-moments,angeli2013:charge-radii}, and just
above the upper end of the statistical uncertainties on the result
$Q/r_p^2=-0.59(3)$ obtained from the GFMC AV18+IL7 calculated quadrupole moment
and radius~\cite{pastore2013:qmc-em-alt9}.

However, taking the predictions for the quadrupole moment by itself, the GFMC
results underpredict experiment (in magnitude) by $\approx24\%$.  In contrast,
NCCI calculations with the Daejeon16 interaction
[Fig.~\ref{fig:q-norm-rp-scan-9be}(a)] show strong indications of convergence
(in particular, shouldering), as well as a clear crossing point for curves of
successive $\Nmax$, suggesting a quadrupole moment clearly larger in magnitude
than for the GFMC AV18+1L7 results and roughly consistent with the experimental
result.  The simple three-point extrapolations (small circles, dotted lines) are
again robust and likewise suggest a value for $Q$ much closer to experiment,
underpredicting experiment (in magnitude) by $\lesssim3\%$, while the
extrapolated radius is consistent with experiment
[Fig.~\ref{fig:q-norm-rp-scan-9be}(d)].

Only for $\isotope[7]{Be}$, among the $p$-shell nuclides, is the charge radius
measured but quadrupole moment unknown.  The calculated $Q/r_p^2$
[Fig.~\ref{fig:q-norm-rp-ratio-teardrop-mirror}(a)] are rapidly converging with
$\Nmax$ and apparently consistent across interactions, with
$Q/r_p^2\approx-1.0$. The measured charge radius~\cite{angeli2013:charge-radii}
yields a proton radius of $r_p=2.52(2)\,\fm$, allowing us to estimate
$Q_p\approx-6.3\,\fm^2$.  However, the apparent systematic discrepancy of the
predicted $Q/r_p^2$ from experiment for the mirror nuclide $\isotope[7]{Li}$
[Fig.~\ref{fig:q-norm-rp-ratio-teardrop-mirror}(g)], noted above, might give us
pause.  In any event, an estimate for the quadrupole moment of $\isotope[7]{Be}$
is already available by calibration to another $E2$ observable, the quadrupole
moment of $\isotope[7]{Li}$, by considering \textit{ab initio} calculations for
the ratio of quadrupole moments for these mirror nuclei, as discussed in
Ref.~\cite{caprio2021:emratio}, where an estimate $Q_p\approx-6.8\,\fm^2$ was
obtained for the unmeasured $\isotope[7]{Be}$ quadrupole moment.

We alternatively use the dimensionless ratio to estimate an unmeasured radius
from a measured quadrupole moment taking, for example, $\isotope[8]{B}$ [Fig.~\ref{fig:q-norm-rp-ratio-teardrop-mirror}(b)] and  $\isotope[12]{B}$
[Fig.~\ref{fig:q-norm-rp-ratio-teardrop-mirror}(l)].  The detailed convergence
properties, for calculations with the Daejeon16 interaction, are illustrated in
Fig.~\ref{fig:q-norm-rp-scan-8b-12b}.

For $\isotope[8]{B}$, the calculated quadrupole moment [Fig.~\ref{fig:q-norm-rp-scan-8b-12b}(a)] is poorly
converged but increasing, with $\Nmax$, in the direction of the experimental
value $Q(2^+)=+6.43(14)\,\fm^2$~\cite{stone2016:e2-moments}.  Similarly,
extrapolations (small symbols, dotted lines) are scattered, but generally
consistent with the experimental value.  The calculated radius
[Fig.~\ref{fig:q-norm-rp-scan-8b-12b}(c)], while showing indications of shouldering,
and exhibiting a crossing point (for curves of successive $\Nmax$) at $\hw\gtrsim10\,\MeV$,
leaves the radius poorly constrained, though apparently $>2.6\,\fm$
(extrapolations suggest $\approx2.7\,\fm$).  Rather than taking the ratio
$Q/r_p^2$ considered thus far, we take the reciprocal and square root to obtain
$r_p/Q^{1/2}$ [Fig.~\ref{fig:q-norm-rp-scan-8b-12b}(e)], from which an estimate for
the radius may be read off directly by calibration to the measured quadrupole
moment (scale at right).  Note that the curves are flattened and compressed,
especially in the range $10\,\MeV\leq\hw\leq20\,\MeV$, allowing us to read off
an estimate of $r\approx2.7\,\fm$.  However, as seen in
Fig.~\ref{fig:q-norm-rp-ratio-teardrop-mirror}(b), the prediction for $Q/r_p^2$
varies by $\approx10\%$ among the interactions considered.  The estimates of $r$
obtained with these different interactions correspondingly vary, by about half
as much ($\approx5\%$).

We may attempt a similar analysis for $\isotope[12]{B}$, where again the
quadrupole moment is measured and the radius unmeasured.  Here the experimental
quadrupole moment is nearly a factor of $5$ smaller than for $\isotope[8]{B}$,
with $Q(1^+)=+1.32(3)\,\fm^2$~\cite{stone2016:e2-moments}.  However, the
rationale for estimating the radius from the measured quadrupole moment is
undermined by the observation that the Daejeon16 interaction appears to be
significantly overestimating the quadrupole moment itself
[Fig.~\ref{fig:q-norm-rp-scan-8b-12b}(b)], even based on the unconverged
calculations so far.  Directly calculating the radius suggests
$r_p\approx2.2\text{--}2.4\,\fm$ [Fig.~\ref{fig:q-norm-rp-scan-8b-12b}(d)], while
normalizing the relatively well-converged ratio $Q/r_p^2$ to the
experimental $Q$ provides a lower estimate, of $r_p\approx2.0\,\fm$
[Fig.~\ref{fig:q-norm-rp-scan-8b-12b}(f)].  Here again, the predicted ratio has a
significant interaction dependence, varying by $\approx1/3$ over the
interactions considered [Fig.~\ref{fig:q-norm-rp-ratio-teardrop-mirror}(l)].
 \begin{figure}
\centering
\includegraphics[width=\ifproofpre{1.0}{0.5}\hsize]{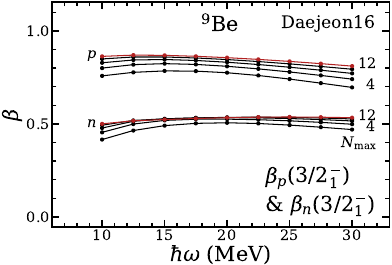}
\caption{Calculated deformations for the $3/2^-$ ground state of
  $\isotope[9]{Be}$, as deduced from the ratios of the form $Q/r^2$, under the
  assumption of axially symmetric rotation.  Shown are $\beta_p$ and $\beta_n$,
  for the proton and neutron distributions, respectively (as labeled).  Results
  are obtained with the Daejeon16 interaction.  Calculated values are shown as
  functions of the basis parameter $\hw$, for successive even values of $\Nmax$,
  from $\Nmax=4$ to $12$ (as labeled).  }
\label{fig:beta-from-ratio-q-rsqr-scan-9be}
\end{figure}

\begin{figure*}
\centering
\includegraphics[width=\ifproofpre{0.70}{1.0}\hsize]{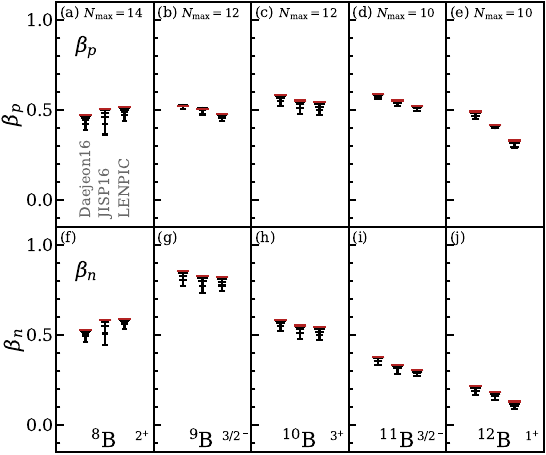}
\caption{Calculated deformations for the ground states of the $p$-shell
  $\isotope{B}$ isotopes ($N=3\text{--}7$, left to right), as deduced from the
  ratios of the form $Q/r^2$, under the assumption of axially symmetric
  rotation.  Shown are $\beta_p$~(top) and $\beta_n$~(bottom), for the proton
  and neutron distributions, respectively.  Results are obtained with the
  Daejeon16, JISP16, and LENPIC interactions (from left to right, within each
  panel).  Calculated values are shown at fixed $\hw=20\,\MeV$ and varying
  $\Nmax$ (increasing tick size), from $\Nmax=4$ to the maximum value indicated
  (at top).  In each case it is assumed that $K=J$, where the ground state
  quantum numbers are indicated for reference (at bottom).}
\label{fig:beta-from-ratio-q-rsqr-teardrop-b}
\end{figure*}

\section{Deformation}
\label{sec:results-moments:deformation}

Finally, let us explore what the ratios of the form $Q/r^2$ indicate for the
ground state quadrupole deformation, via the rotational
relation~(\ref{eqn:Q-norm-r-beta-proton}).  We first illustrate with the
$\isotope[9]{Be}$ ground state, in
Fig.~\ref{fig:beta-from-ratio-q-rsqr-scan-9be}, which we take to be the band
head of a rotational band with
$K=3/2$~\cite{millener2001:light-nuclei,millener2007:p-shell-hypernuclei,caprio2020:bebands}.
In addition to the ratio $Q/r_p^2$ of proton observables (or, for consistency of
notation across protons and neutrons, $Q_p/r_p^2$), which yields $\beta_p$
(upper curves), we consider the corresponding ratio $Q_n/r_n^2$ of neutron
observables, which yields $\beta_n$ (lower curves).  The curves for $\beta_p$
show the same underlying dimensionless ratio as in
Fig.~\ref{fig:q-norm-rp-scan-9be}(j), simply rescaled according
to~(\ref{eqn:Q-norm-r-beta-proton}).

Note the much smaller calculated deformation for the neutrons, by nearly a
factor of $2$.  Such is to be expected in a cluster molecular orbital
interpretation~\cite{okabe1979:9be-molecular-part-1,vonoertzen1996:be-molecular,kanadaenyo1999:10be-amd},
in which $\isotope[9]{Be}$ is an $\alpha$-$\alpha$ dimer plus a covalent
neutron, when the neutron occupies an equatorial ($\pi$) orbital about the
molecular symmetry axis, leading to a more spherical shape.  This cluster
interpretation is supported by densities obtained with NCCI calculations (see
Fig.~7 of Ref.~\cite{maris2013:ncsm-pshell}).

Turning to the boron isotopes, shown in
Fig.~\ref{fig:beta-from-ratio-q-rsqr-teardrop-b}, the ground state angular
momenta for all these isotopes support a quadrupole moment
($J\geq1$). For each isotope, the deformation is extracted from the quadrupole
moment and radius, on the assumption that the ground state is a rotational band
head, that is, taking $K=J$ in~(\ref{eqn:Q-norm-r-beta-proton}).\footnote{Within a
  rotational picture, a $J=3/2$ ground state could alternatively be a member of a $K=1/2$
  band, if the energies are inverted due to Coriolis decoupling, as in
  $\isotope[7]{Be}$~\cite{millener2007:p-shell-hypernuclei,caprio2020:bebands}.
  Furthermore, a $J=1$ ground state could alternatively be a member of a $K=0$ band with odd
  signature~\cite{rowe2010:collective-motion}.  Nonetheless, the calculated
  low-energy spectra for $\isotope[9,11,12]{B}$ are consistent with a simple $K=J$
  assignment, as is also expected for these nuclei from $\grpsu{3}$ symmetry
  arguments~\cite{millener2007:p-shell-hypernuclei}.}  We again consider both
the ratio of proton observables, for $\beta_p$
[Fig.~\ref{fig:beta-from-ratio-q-rsqr-teardrop-b} (top)], and that of neutron
observables, for $\beta_n$ [Fig.~\ref{fig:beta-from-ratio-q-rsqr-teardrop-b}
  (bottom)].  The calculated ratios $Q/r_p^2$ themselves vary by almost an order
of magnitude across these isotopes:
$\isotope[8]{B}$ [Fig.~\ref{fig:q-norm-rp-ratio-teardrop-mirror}(b)],
$\isotope[9]{B}$ [Fig.~\ref{fig:q-norm-rp-ratio-teardrop-mirror}(c)],
$\isotope[10]{B}$ [Fig.~\ref{fig:q-norm-rp-ratio-teardrop-neqz}(b)],
$\isotope[11]{B}$ [Fig.~\ref{fig:q-norm-rp-ratio-teardrop-mirror}(k)],
and
$\isotope[12]{B}$ [Fig.~\ref{fig:q-norm-rp-ratio-teardrop-mirror}(l)].
However, after application of the rotational
relation~(\ref{eqn:Q-norm-r-beta-proton}), the proton deformations are
relatively constant, at $\beta_p\approx0.5$, across the isotopic chain
[Fig.~\ref{fig:beta-from-ratio-q-rsqr-teardrop-b} (top)].  In contrast, the
neutron deformation starts at $\beta_n\approx0.5$ for $\isotope[8]{B}$
[Fig.~\ref{fig:beta-from-ratio-q-rsqr-teardrop-b}(f)], peaks at
$\beta_n\approx0.8$ for $\isotope[9]{B}$
[Fig.~\ref{fig:beta-from-ratio-q-rsqr-teardrop-b}(g)], then steadily decreases
to $\beta_n\approx0.1$--$0.2$ for $\isotope[12]{B}$
[Fig.~\ref{fig:beta-from-ratio-q-rsqr-teardrop-b}(j)].

Such a direct interpretation of the dimensionless ratios in these boron isotopes as
yielding $\beta$ parameters for axially symmetric deformation should, however, be regarded
with caution.  One must keep in mind the possible influence of triaxiality in these
nuclei~\cite{kanadaenyo2001:amd,suhara2010:amd-deformation}, as in the
neighboring beryllium and carbon
isotopes~\cite{kanadaenyo1997:c-amd-pn-decoupling,kanadaenyo1999:10be-amd,kanadaenyo2001:amd,bohlen2007:10be-pickup,caprio2022:10be-shape-sdanca21}.

\section{Conclusion}
\label{sec:concl}

Although \textit{ab initio} predictions of nuclear $E2$ observables are hampered
by poor convergence in NCCI calculations, correlations among calculated
observables can be exploited to extract meaningful predictions.  In particular,
systematic truncation errors can cancel in appropriate dimensionless ratios of
observables, leading to more robust convergence, as we explore here for
dimensionless ratios of the form $Q/r^2$, involving quadrupole moment and radius
observables.

The nuclear ground state charge radius and/or quadrupole moment are
well-measured for many nuclei.  If both are known, an
\textit{ab initio} prediction of the ratio $Q/r_p^2$ provides a test of the
\textit{ab initio} description, including structural features which may be
sensitive to the interaction.  If only the charge radius is well-measured, an
\textit{ab initio} prediction of the ratio $Q/r_p^2$ effectively yields a
prediction for the quadrupole moment itself.  In practice, however, the
quadrupole moment is more often measured, while the radius is still unknown, in
which case, conversely, an \textit{ab initio} prediction for $Q/r_p^2$ yields a
prediction for the unmeasured radius, as illustrated here for $\isotope[8]{B}$
and $\isotope[12]{B}$.  Similar observations apply to neutron quadrupole moment
and radius observables, to the extent that they may be experimentally
accessible.

An approach which seeks to exploit correlations in the convergence of two
observables, in this case through the dimensionless ratio $Q/r^2$, is most
suited to eliminating smoothly varying systematic truncation error.  Such an
approach cannot be expected to anticipate abrupt structural changes arising as a
function of the basis truncation, as when levels undergo an avoided
crossing~\cite{mccoy2024:12be-shape}.  Even in the absence of such a dramatic
scenario, one can readily imagine the convergence of the ratio to be confounded
by any significant role for two-state mixing in the dependence of the calculated
observables on the basis trunction~\cite{casten2000:ns}.

While we have focused here on the ground state quadrupole moment, excited state
quadrupole moments may be treated similarly, \textit{e.g.}, considering the
ratio $Q(2^+)/r_p^2(0^+)$ for an even-even nucleus [see
  Fig.~\ref{part2:fig:be2-norm-rp-scan-12c}(f) of Part~II].  More general
correlations between the convergence properties of radius and quadrupole moment
observables~\cite{knoell:ncsm-ml-extrapolation-em-PREPRINT} may also be
considered.  Continuing the chain of correlations, one may consider calibrating
to the radius in predicting other observables which are known to be correlated
to the quadrupole moment, namely, weak-interaction recoil-order form
factors~\cite{sargsyan2022:8li-clustering-beta-recoil}.

Moreover, \textit{ab initio} calculated ratios of the form $Q/r^2$, for both
proton and neutron observables, provide insight into the nuclear quadrupole
deformation.  In an axial rotational description, these ratios measure the
quadrupole deformation parameter $\beta$.  In calculations for the $p$-shell
$\isotope{B}$ isotopes, we find an essentially constant proton deformation, and
a smooth evolution in neutron deformation peaking just below mid-shell, to the
extent that an axial rotor description applies to these nuclei.

\begin{acknowledgments}
  We thank James P.~Vary, Ik Jae Shin, and Youngman Kim for sharing illuminating
  results on ratios of observables, Greg Hackman and Peter Mueller for valuable
  discussions, and Scott R.~Carmichael and Shwetha L.~Vittal for comments on the
  manuscript.  This material is based upon work supported by the U.S.~Department
  of Energy, Office of Science, under Awards No.~DE-FG02-95ER40934,
  DE-AC02-06CH11357, and DE-SC0023495 (SciDAC5/NUCLEI).  This research used
  resources of the National Energy Research Scientific Computing Center (NERSC),
  a DOE Office of Science User Facility supported by the Office of Science of
  the U.S.~Department of Energy under Contract No.~DE-AC02-05CH11231, using
  NERSC awards NP-ERCAP0020422 and NP-ERCAP0023497.
\end{acknowledgments}

\appendix

\section{Relations for exponential convergence}
\label{sec:app-geometric}

Under the assumption of exponential
convergence, of the form~(\ref{eqn:exp-convergence}),
\begin{equation*}
  X(\Nmax)=X_{\infty}+a\exp(-c\Nmax),
\end{equation*}
it is readily verified,
\textit{e.g.}, from the standard closed form for a geometric
series~\cite{olver2010:handbook}, that the difference between the limiting
value $X_{\infty}$ and the last calculated value $X$ may be related to
the size $\Delta$ of the last
step taken, in terms of the ratio $\eta$, from~(\ref{eqn:difference-ratio-from-delta}), of the sizes of successive steps.
Namely, 
\begin{equation}
   \label{eqn:generalized-zeno-ratio}
    X_\infty-X=\frac{\eta}{1-\eta}\Delta.
\end{equation}

In the familiar case of Zeno's paradox, where each step that one takes is half
the size of the preceding step ($\eta=1/2$), each step brings one half-way to
limiting value (\textit{i.e.}, the ``finish line'').  Consequently, to reach the
limiting value, one must proceed as far again as in the last step taken.
From~(\ref{eqn:generalized-zeno-ratio}), we indeed recover $X_\infty-X=\Delta$.
Then, for example, for $\eta=2/3$, one must proceed twice as far again as in the
last step taken; for $\eta=3/4$, one must proceed $3$ times as far; for
$\eta=4/5$, one must proceed $4$ times as far, \textit{etc.}

Given three consecutive calculated values, namely, $X(\Nmax-4)$, $X(\Nmax-2)$,
and $X(\Nmax)$, an exponentential fit~(\ref{eqn:exp-convergence}) to these three
points is exact.  From this same observation~(\ref{eqn:generalized-zeno-ratio}),
we may extract the value of $X_{\infty}$ corresponding to this exact fit
analytically (that is, rather than by numerical least-squares fitting) simply by
evaluating
\begin{equation}
  \label{eqn:three-point-extrapolation}
  X_\infty = X(\Nmax) + \frac{\eta(\Nmax)}{1-\eta(\Nmax)}\Delta(\Nmax),
\end{equation}
where we recall that the ratio $\eta(\Nmax)$ may be evaluated,
by~(\ref{eqn:difference-ratio-from-delta}), in terms of the differences
$\Delta(\Nmax)$ and $\Delta(\Nmax-2)$, and thus in terms of $X(\Nmax)$, $X(\Nmax-2)$,
and $X(\Nmax-4)$.  For simplicity of notation in
the above expressions, we have not explicitly labeled $\Delta$ or $\eta$ by the
observable $X$ to which they correspond (as they were in
Sec.~\ref{sec:results-moments:9be}).

\bibliographystyle{apsrev4-2}
\nocite{control:title-on}

\end{document}